\documentclass{aa}
\usepackage{verbatim} 
\usepackage{amsmath}
\usepackage{amssymb}
\usepackage{graphicx}
\usepackage{booktabs}
\usepackage{natbib}
\usepackage[colorlinks=true,linkcolor=blue,urlcolor=blue,citecolor=blue]{hyperref}
\usepackage{xspace}
\usepackage{xcolor}
\usepackage{placeins}

\makeatletter

\usepackage{txfonts}\usepackage{twoopt}
\bibpunct{(}{)}{;}{a}{}{,}
\usepackage{enumitem}
\defcitealias{paperII}{Paper~II}

\newcommand{\msun}{{\rm M}_{\odot}}
\newcommand{\lsun}{{\rm L}_{\odot}}
\newcommand{\rsun}{{\rm R}_{\odot}}

\newcommand{\MESA}{\texttt{MESA}\@\xspace}
\newcommand{\qty}[2]{\ensuremath{#1\,\mathrm{#2}}}

\newcommand{\eg}{e.g.\@\xspace}
\newcommand{\cf}{c.f.\@\xspace}
\newcommand{\ie}{i.e.\@\xspace}
\titlerunning{Supernova light curves from pulsating stars}
\authorrunning{V.A.~Bronner}
\makeatother
\begin{document}
\title{Explosions of pulsating red supergiants: a natural pathway for the diversity of Type~II-P/L supernovae}
\author{V.A.~Bronner\inst{\ref{HITS},\ref{UniHD}}\thanks{vincent.bronner@h-its.org}, E.~Laplace\inst{\ref{HITS},\ref{KUL},\ref{KULG},\ref{API}}, F.R.N.~Schneider\inst{\ref{HITS},\ref{ZAH}}, Ph.~Podsiadlowski\inst{\ref{LISA},\ref{HITS},\ref{Oxford}}}
\institute{
Heidelberger Institut f\"{u}r Theoretische Studien, Schloss-Wolfsbrunnenweg 35, 69118 Heidelberg, Germany\label{HITS}
\and
Universit\"{a}t Heidelberg, Department of Physics and Astronomy, Im Neuenheimer Feld 226, 69120 Heidelberg, Germany\label{UniHD}
\and
Institute of Astronomy, KU Leuven, Celestijnenlaan 200D, B-3001 Leuven, Belgium\label{KUL}
\and
Leuven Gravity Institute, KU Leuven, Celestijnenlaan 200D, box 2415, 3001 Leuven, Belgium\label{KULG}
\and
Anton Pannekoek Institute of Astronomy, University of Amsterdam, Science Park 904, 1098 XH Amsterdam, The Netherlands\label{API}
\and
Zentrum f\"{u}r Astronomie der Universit\"{a}t Heidelberg, Astronomisches Rechen-Institut, M\"{o}nchhofstr. 12-14, 69120 Heidelberg, Germany\label{ZAH}
\and
London Centre for Stellar Astrophysics, Vauxhall, London\label{LISA}
\and
University of Oxford, St Edmund Hall, Oxford OX1 4AR, UK\label{Oxford}
}
\date{Received xxx / Accepted yyy}
\abstract{
    Red supergiants (RSGs), which are progenitors of hydrogen-rich Type~II supernovae (SNe), have been known to pulsate from both observations and theory. The pulsations can be present at core collapse and affect the resulting SN. However, SN light curve models of such RSGs commonly use hydrostatic progenitor models and ignore pulsations. Here, we model the final stages of a \qty{15}{\msun} RSG and self-consistently follow the hydrodynamical evolution. We find the growth of large amplitude radial pulsations in the envelope. After a transient phase where the envelope restructures, the pulsations settle to a steady and periodic oscillation with a period of \qty{817}{days}. We show that they are driven by the $\kappa\gamma$-mechanism, which is an interplay between changing opacities and the release of recombination energy of hydrogen and helium. This leads to complex and non-coherent expansion and contraction in different parts of the envelope, which greatly affect the SN progenitor properties, including its location in the Hertzsprung-Russell diagram. We simulate SN explosions of this model at different pulsations phases. Explosions in the compressed state result in a flat light curve (Type~II-P). In contrast, the SN light curve in the expanded state declines rapidly, reminiscent of a Type~II-L SN. For cases in between, we find light curves with various decline rates. Features in the SN light curves are directly connected to features in the density profiles. These are in turn linked to the envelope ionization structure, which is the driving mechanism of the pulsations. We predict that some of the observed diversity in Type~II SN light curves can be explained by RSG pulsations. For more massive RSGs, we expect stronger pulsations that might even lead to dynamical mass ejections of the envelope and to an increased diversity in SN light curves.
}
\maketitle

\section{Introduction\label{sec:introduction}}

Red supergiants (RSGs) mark the end stages of stellar evolution in a certain mass range of massive stars. Some defining features of RSGs are that they are located at the cool edge of the Hertzsprung-Russell diagram (HRD) and have a deep convective and hydrogen-rich envelope that extends from several hundred to thousands of solar radii in size. Observationally, RSGs are known to show (periodic) brightness variations \citep{kiss2006a,percy2014a} -- some well-known variable RSGs are Betelgeuse ($\alpha$~Ori), VY~CMa, TV~Gem, VX~Sgr, and S~Per. These brightness variations have been attributed to radial pulsations in the envelope of the RSGs \citep{stothers1969a,stothers1971a,heger1997a,wood1983a,guo2002a,yoon2010a} and are thought to be driven by the $\kappa$-mechanism. Alternative explanations for the irregular variability include the brightness variations caused by the large convective cells present in the envelopes of RSGs \citep{schwarzschild1975a,antia1984a}. 

Most stars that end their lives as RSGs explode in a supernova (SN).
Supernovae that show a clear signature of hydrogen are classified as Type~II SNe \citep{minkowski1941a} and are connected to explosions of massive stars with a hydrogen-rich envelope \citep{smartt2009a,smartt2015a}. Type~II-P SNe are a subclass of Type~II SNe whose light curves show a prominent plateau phase of $\approx \qty{100}{d}$. Type~II-L SNe are a second class of hydrogen-rich SNe with a linearly declining light curve \citep{barbon1979a}. Because of the low numbers of Type~II-L SN observations \citep{smartt2009a}, it was believed that there is a clear distinction between II-P and II-L SNe \citep{patat1993a,patat1994a,arcavi2012a}. However, as the numbers of SN observations increased, there seemed to be a less clear distinction between these two categories and more of a continuum between II-P and II-L SNe, characterized by different decline rates of the light curve during the first $\approx \qty{100}{d}$ \citep{anderson2014a,faran2014a,faran2014b,sanders2015a}. Additionally, supernovae with a fast-declining light curve are found to be more luminous compared to supernovae with lower decline rates \cite{anderson2024a}. 

Connecting supernova properties and the light-curve decline rate to their exact progenitors remains a challenge, though significant progress has been made in recent years. Using archival imaging, it is possible to identify RSG as the direct progenitor stars of the classical Type~II-P SNe that have low decline rates \citep{smartt2009a}. The progenitors of SNe with high decline rate, \ie, classically referred to as Type~II-L SNe, are not well known due to their scarcity. There seems to be a trend that yellow supergiants might be possible progenitors \citep{vandyk2016a}. However, it should be noted that there are exceptions to this trend and \citet{valenti2016a} claim that there is no clear distinction between the progenitors of SNe with faster or slower decline rates.

Numerical models computing the SN light curves from pre-SN stellar structures show that the explosions of RSGs are indeed able to reproduce observed light curves of hydrogen-rich SNe with flat light curves \citep[\eg][]{litvinova1983a,chieffi2003a,bersten2011a,dessart2013a,morozova2015a}. Producing fast-declining SNe from numerical models can be achieved by stellar structure with less massive hydrogen-rich envelopes \citep{grassberg1971a,young1989a,blinnikov1993a,schlegel1996a}, although reducing the envelope mass alone fails to reproduce the observed luminosities \citep[\eg][]{hillier2019a}. Potential sources for reduced envelope masses can be strong mass loss in the RSG phase \citep{moriya2011a,georgy2012a,fuller2024a}, mass transfer in binary systems \citep{podsiadlowski1992a,nomoto1995a,dessart2024b}, or even eruptive mass loss shortly before the SN \citep{smith2014b,clayton2018a}. The presence of strong mass loss is backed up observationally, as many SNe with fast-declining light curves have a larger peak brightness and often show signs of interaction with circumstellar material (CSM) \citep{smith2014a,bostroem2019a,zhang2024a,jacobson-galan2024b}.

Supernova light curve calculations require an input stellar structure that usually comes from a one-dimensional (1D) stellar evolution calculation. Generally, hydrostatic equilibrium is assumed in such stellar evolution calculations. This is also the case for pre-SN structures that enter SN light curve calculations.\footnote{Hydrostatic equilibrium might only be assumed in the envelope. The core is collapsing and therefore not in hydrostatic equilibrium.} Large-scale radial pulsations in RSGs are usually ignored. However, the envelope structure of a pulsating RSG might significantly deviate from a hydrostatic envelope, depending on the amplitudes, but the structure of the hydrogen-rich envelope is critical in determining the shape of the resulting SN light curve. One notable exception is the work by \citet{goldberg2020a}, which studies the effect of the pulsations of RSGs just before the onset of core collapse. They analyzed at which frequencies the RSGs would pulsate and then trigger these pulsations just before the SN via velocity perturbations in the envelope. Here, we aim to simulate the pulsations self-consistently whenever the envelopes of RSGs become dynamically unstable.

In this work, we study the radial pulsations in the envelopes of RSGs, their impact on the envelope structure, and the light curves of the resulting SNe. First, we describe the computational tool we use in Sect.~\ref{sec:methods}. This includes a 1D stellar evolution code to study the pulsations, a dust radiation code to predict light curves of the pulsating RSGs, a semi-analytic code to predict the core-collapse outcome, and a one-dimensional hydrodynamic supernova code to compute the SN light curves. In Sect.~\ref{sec:pulsations}, we analyze in detail the pulsations of RSGs, their origin, their driving mechanism, and their effects on the envelope structure. Then, we model the explosion of the pulsating RSG and compute the resulting SN light curve in Sect.~\ref{sec:LC}. In Sect.~\ref{sec:discussion}, we discuss some uncertainties of our models and the implications for SN light curve modeling, before summarizing and concluding in Sect.~\ref{sec:conclusion}.

\section{Methods}\label{sec:methods}

In this section, we describe the methods used to compute the pulsating RSG and the SN light curves. The stellar-evolution models are described in Sect.~\ref{sec:methods:SE}. Sect.~\ref{sec:methods:pulsations} summarizes the methods used to model the pulsations of the RSG. The dust computations are explained in Sect.\ref{sec:methods:dust}. Lastly, we describe the SN light curve modeling in Sect.~\ref{sec:methods:LC}.

\subsection{Stellar evolution models}\label{sec:methods:SE}
We evolve an initially \qty{15}{\msun} stellar model from the zero-age main sequence all the way to core collapse, using the Modules for Experiments in Stellar Astrophysics (\MESA) stellar-evolution code revision 10398 \citep{paxton2011a,paxton2013a,paxton2015a,paxton2018a,paxton2019a,jermyn2023a} with the same setup as in \citet{schneider2021a} and \citet{laplace2025a}. In particular, we use a mixing-length parameter $\alpha_\mathrm{MLT}=1.8$ and employ the Ledoux criterion to model convection. We use MLT++ for better numerical stability because it increases the efficiency of energy transport in convective regions that are close to the Eddington luminosity. Convective-boundary mixing is implemented using step-overshooting of 0.2 pressure-scale heights on top of convective core hydrogen and helium burning. There is no convective-boundary mixing in all other convective shells. Additionally, we use semi-convection with an efficiency factor of $\alpha_\mathrm{sc}=0.1$.

The model is evolved to core-collapse, defined when iron-core infall velocity exceeds \qty{950}{km\,s^{-1}}. At the end of core carbon burning, defined as the moment when the core carbon mass fraction is lower than $10^{-4}$, the star appears as a RSG with a mass of \qty{12.2}{\msun}, a radius of \qty{1024}{\rsun}, a luminosity of \qty{1.12\times 10^5}{\lsun} and an effective temperature of \qty{3304}{K}. The bottom of the convective envelope is at a mass coordinate of \qty{5.23}{\msun}. This is the starting point for the subsequent dynamical modeling of the RSG.

\subsection{Modelling of the pulsations}\label{sec:methods:pulsations}
We use the model of the \qty{15}{\msun} RSG at the end of core carbon burning as the starting point for the dynamical calculations. First, we enforce hydrostatic equilibrium throughout the entire model. The default settings of \MESA for massive stars, as found in \texttt{inlist\_massive\_defaults}, lift the assumption of hydrostatic equilibrium in layers exceeding a temperature of \qty{10^8}{K} when the time step is smaller than \qty{0.1}{yr}. This way, the contraction of the core during the final burning stages and the onset of core collapse can be directly modeled. The temperature cut-off ensures that the hydrodynamical mode is enabled only in the core and not in the envelope. After forcing the model into hydrostatic equilibrium, we excise the core at a mass coordinate of \qty{4.6}{\msun}. This is about \qty{90}{\%} of the helium core mass when defining the helium core by a hydrogen abundance of less than 0.1. The exact mass coordinate at which we remove the core has a negligible effect on the subsequent simulations, as long as it is within the helium core and outside the helium-burning shell (see Appendix~\ref{app:Mcut}). It is necessary to enforce hydrostatic equilibrium before excising the core to have a static inner boundary condition that does not have any long-term effect on the dynamical calculations.

Once the velocity is forced to zero throughout the star and the core is removed, we enable the hydrodynamic modeling in \MESA throughout the entire simulated domain. We use a basic setup for the hydrodynamic modeling that is similar to the setup used in \citet{clayton2017a} and \citet{bronner2024a}. We use \MESA's implicit hydrodynamic capabilities as described in \citet{paxton2015a} that make use of artificial viscosity. For the outer boundary condition, we use the default setting from \MESA, that is, modeling an atmosphere using the Eddington $T\mbox{--}\tau$ relation \citep{eddington1926a,paxton2013a}. To ensure that we can resolve the dynamical evolution in the envelope, we limit the time steps to \qty{10^{-2}}{yr}. Finally, we turn off any change in abundances from nuclear burning while still keeping the energy generation rate. This keeps the star in a chemically stable configuration without any nuclear timescale evolution and is justified because of the relatively long timescale of the chemical evolution compared to the dynamical evolution of the envelope.

We end the simulation after \qty{300}{yr}, which is much longer than the initial thermal timescale of the envelope of about \qty{11}{yr}. Longer simulations do not introduce any long-term changes because we disable the conversion of elements via nuclear burning, and we use a static inner boundary condition.

There are several limitations to our pulsation model, most of which originate from the 1D nature of the model and the associated lack of 3D effects on the pulsations. These limitations are presented and discussed in detail in Sect.~\ref{sec:discussion:lims}. For the models of the pulsating RSG, we in particular keep using MLT++ and a mixing-length parameter $\alpha_\mathrm{MLT}=1.8$ for the envelope. There exists work suggesting increasing the value of $\alpha_\mathrm{MLT}$ in the envelope of RSGs to produce SNe that align better with observations \citep[\eg,][]{dessart2013a,paxton2018a,goldberg2022a}. This is also addressed in Sect.~\ref{sec:discussion:lims}.

\subsection{Dust modeling}\label{sec:methods:dust}
Some RSGs are believed to be surrounded by a dust shell \citep{verhoelst2009a}. Such a dust shell might originate from the mass loss via winds and could be enhanced by pulsations and eruptive mass loss \citep{bonanos2024a,dewit2024a}. We use the dust radiation-transport code \texttt{DUSTY} \citep{ivezic1997a,ivezic1999a} to compute light curves in different broad-band filters of the RSG model, assuming it is surrounded by a dust shell. For the dust shell, we use common assumptions for RSGs. The dust is composed of warm silicates \citep{ossenkopf1992a}, and the grain-size distribution follows a power law with exponent $-3.5$ and minimum and maximum grain size of \qty{0.005}{\mu m} and \qty{0.25}{\mu m} \citep{mathis1977a}. The dust temperature at the inner boundary $r_1$ is set to \qty{800}{K} and the density distribution of the dust shell follows a power law with exponent $-2$. For the total extent of the dust shell, we use a relative thickness of $r_\mathrm{out}/r_1 = 1000$. The total optical depth $\tau$, defined at \qty{0.55}{\mu m}, is treated as a free parameter and is varied between 0.1 and 100 for each \texttt{DUSTY} model. These assumptions might not be valid for a post-core-helium burning RSG that we consider. Convective mixing after core helium burning could have changed the envelope composition and hence the dust chemistry. Additionally, some RSGs are surrounded by exceptionally large dust grains \citep{scicluna2015a}. Because we are mostly interested in the qualitative effects of a dust shell around a pulsating RSG, we refrain from a more detailed analysis of the dust parameters, but study this in \citet{paperII}, hereafter \citetalias{paperII}.

One main input for the dust radiation calculations is the spectral shape of the illuminating radiation source (\ie, the pulsating RSG). We use the MARCS models \citep{gustafsson2008a} to describe the spectrum of the RSG. We select the subset of spherical \qty{5}{\msun} models\footnote{There also exists a smaller grid of \qty{15}{\msun} MARCS models that covers a narrower range of effective temperatures. We find excellent agreement between the 5 and \qty{15}{\msun} models in the parameter range that both grids cover. Additionally, an even denser \qty{1}{\msun} grid is available that also agrees well with the \qty{5}{\msun} models. Contrarily, when using a black-body spectrum as the input radiation source, we find deviations of up to \qty{2}{mag} compared to the MARCS models.} with standard abundances, solar metallicity, and a microturbulence parameter of 5. This subset of MARCS models is available on a well-defined grid of $T_\mathrm{eff}$ and $\log g$.

We compute \texttt{DUSTY}+MARCS (DM) models as described above. For each DM model, we compute the absolute $V$- and $K$-band magnitudes\footnote{All magnitudes are reported in the Vega system \citep{johnson1966a}.} from the dust-attenuated spectral energy distribution. To construct light curves of the RSG, we linearly interpolate between the DM model based on the $T_\mathrm{eff}$ and $\log g$ of the RSG (for more details on the interpolation, see Appendix~\ref{app:DM_interp}). The DM models only cover temperatures of $T_\mathrm{eff} > \qty{2400}{K}$, such that we are unable to predict the parts of the light curve where the RSG is cooler than \qty{2400}{K}. For temperatures cooler than \qty{2400}{K}, we do not make any predictions, because extrapolation may introduce large uncertainties. The final free parameter of the light curves is the total optical depth $\tau_0$ at the beginning of the pulsation cycle. This parameter corresponds to the total dust mass around the RSG and cannot be determined a priori. The total optical depth $\tau_0$ can be converted to the total dust mass. This corresponds to dust masses between $10^{-4}$ and $10^{-2}\,\msun$ for $\tau_0 = 0.5-10$, assuming a dust grain density of $3 \, \mathrm{g\, cm^{-3}}$.  We do not assume any extinction and reddening for these theoretical predictions.

\subsection{Supernova light curve models}\label{sec:methods:LC}
To determine the supernova explosion parameters of the RSG, we employ the semi-analytic supernova code by \citet{muller2016a} with the same calibrations as \citet{schneider2021a} and \citet{temaj2024a}. Based on our stellar model at the onset of core collapse, applying the code results in an explosion energy of $E_{\rm{exp}} = \qty{1.60}{B}$, a mass of $M_{\rm{NS}} = \qty{1.502}{\msun}$ for the compact object remnant, and a mass $M_{\rm{Ni}} = \qty{0.086}{\msun}$ of synthesized $^{56}\rm{Ni}$. We employ these explosion parameters to compute the supernova light curve with the 1D open-source Lagrangian radiation hydrodynamics code \texttt{SNEC} \citep{morozova2015a} version 1.0.1. \texttt{SNEC} assumes Local Thermodynamic Equilibrium (LTE) for computing the light curves, which is a reasonable assumption during the plateau phase but is less accurate for the shock breakout and nebular phases, which are known to be affected by non-thermal effects. To compute the SN light curves of our progenitor model at different pulsation phases, we map each stellar structure on a Lagrangian grid containing 1000 points. We use the thermal bomb model implemented in \texttt{SNEC} to compute the propagation of a shock through the stellar structure. After excising the mass $M_{\rm{NS}}$ from the grid, we inject enough energy in the \qty{0.1}{\msun} above this mass coordinate to reach the desired final energy $E_{\rm{exp}}$. We simulate strong mixing of nickel in the ejecta by spreading $M_{\rm{Ni}}$ of nickel from the inner boundary to 90\% of the final mass. We keep the other properties, such as the opacity, equation of state, and ionization states, to the default values for II-P SNe in \texttt{SNEC} and follow the shock propagation until about \qty{200}{d} after the explosion.

\section{Pulsations of red supergiants}\label{sec:pulsations}

In this section, we present the results of the pulsation calculations. First, we show the growth of the pulsations in Sect.~\ref{sec:pulsations:growth} and describe the pulsation mechanism in Sect.~\ref{sec:pulsations:mechanism}. Then, we show how the pulsations change the density structure of the RSG in Sect.~\ref{sec:pulsations:density}, and how such a pulsating RSG might appear observationally by applying a dust shell model around the RSG and computing photometric light curves in Sect.~\ref{sec:pulsations:dusty}.

\subsection{Growth of pulsations and envelope restructuring}\label{sec:pulsations:growth}

\begin{figure*}
    \centering
    \includegraphics[width=17cm]{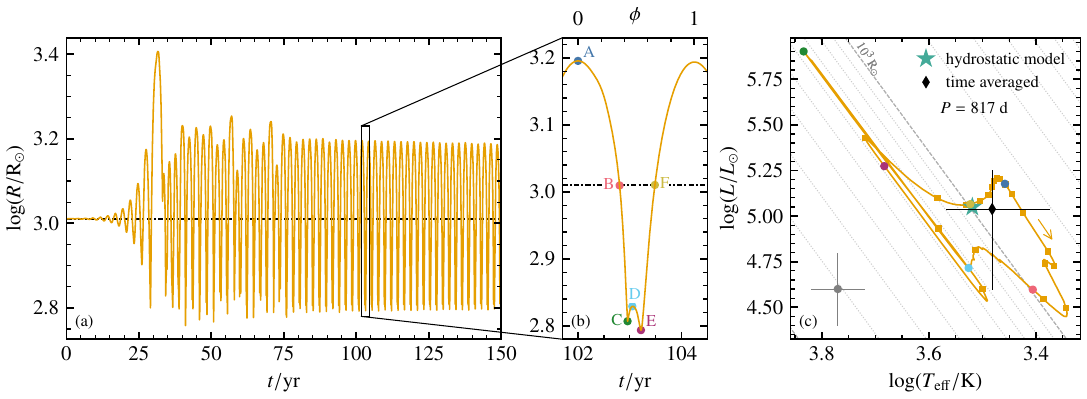}
    \caption{Radial pulsation of the RSG. The radius evolution is shown in panel (a) with a zoom-in on one pulsation cycle in panel (b). The black dash-dotted line indicates the radius of the hydrostatic model. Panel (c) shows one pulsation cycle in the Hertzsprung-Russell diagram (HRD). The arrow indicates the direction of the loop in the HRD, and the markers are spaced equally in time every $1/20$ of the pulsation period. The hydrostatic model, the time-averaged effective temperature and luminosity, and their uncertainties are shown as well. For comparison, we show the typical uncertainty of the luminosity and effective temperature of Type~II SN progenitors from \citet{smartt2015a} as a gray marker. The colored markers labeled A--F in panels (b) and (c) indicate characteristic stages during the pulsation at which we calculate SN light curves. The pulsation phase $\phi$ is defined to be zero at maximum expansion.}
    \label{fig:pulsations_HRD}
\end{figure*}

\begin{figure}
    \centering
    \resizebox{\hsize}{!}{\includegraphics{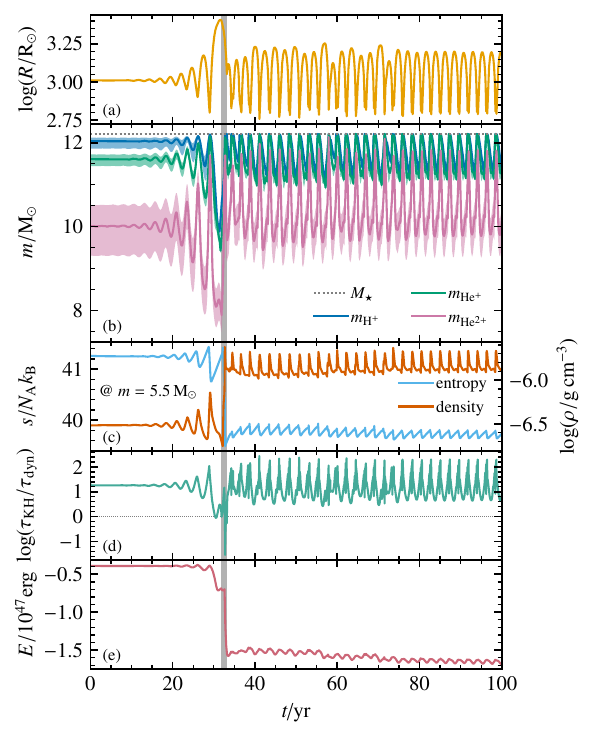}}
    \caption{Restructuring of the envelope during the initial transient. Panel~(a) shows the radius of the envelope (similar to Fig.~\ref{fig:pulsations_HRD}a). The partial ionization zones and the mass coordinates of the recombination fronts of hydrogen and helium are shown in panel~(b), as well as the total mass of the star. The recombination fronts are defined where the ionization fraction of the respective species reaches 0.5. The partial ionization zones are defined to be the layers where the ionization fraction of the respective species is between 10 and \qty{90}{\%}. Panel~(c) shows the specific entropy in units of Avogadro's number $N_\mathrm{A}$ and the Boltzmann constant $k_\mathrm{B}$, as well as the density. Both quantities are reported at a mass coordinate of \qty{5.5}{\msun}, which is close to the bottom of the convective envelope. Panel (d) shows the radiative cooling timescale in units of the dynamical timescale. The total energy of the convective envelope is shown in panel (e). The gray band indicates the catastrophic cooling event at $t \approx \qty{32.5}{yr}$.}
    \label{fig:restructuring}
\end{figure}

Once we turn on the hydrodynamic mode of \MESA at the end of core carbon burning, the RSG naturally starts to pulsate radially (Fig.~\ref{fig:pulsations_HRD}). The amplitude of the radial pulsations grows exponentially for $t \lesssim \qty{30}{yr}$. We compute the growth rate $\eta$ as defined in \citet{yoon2010a} via $\eta = |v(t_0 + P)/v(t_0)|$, where $v(t_0)$ is a relative maximum of the surface velocity and $P \approx \qty{950}{d}$ is pulsation period. We find $\eta \approx 1.87$. This is lower than the value of $\eta = 2.15$ from \citet[their equation~1]{yoon2010a}, but still within their observed spread \citep[\cf][Fig.~3]{yoon2010a}.

\begin{figure}
    \centering
    \resizebox{\hsize}{!}{\includegraphics{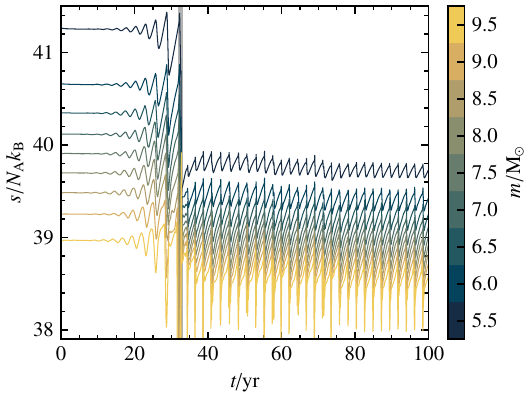}}
    \caption{Specific entropy $s$ in units of Avogadro's number $N_\mathrm{A}$ and the Boltzmann constant $k_\mathrm{B}$ at various mass coordinates inside the convective envelope. The gray band indicates the restructuring event at $t \approx \qty{32.5}{yr}$.}
    \label{fig:restructuring_entropy}
\end{figure}

At a time of around \qty{32.5}{yr}, there are drastic changes within the entire envelope that cause changes in both the pulsation amplitude and period (\ie, just after peak radius in Figs.~\ref{fig:pulsations_HRD}a and~\ref{fig:restructuring}a). Figure~\ref{fig:restructuring}c shows that the density near the bottom of the convective envelope increases. At the same time, the entropy decreases. This change is caused by a catastrophic cooling event at the surface of the extended envelope. Initially, the pulsation amplitude grows exponentially, oscillating around the equilibrium radius. At the same time, the partial ionization zones of hydrogen and helium periodically move inward and outward (Fig.~\ref{fig:restructuring}b). This means that ionization energy is periodically released (\ie, usually referred to as recombination energy) during the expansion, and energy from the pulsations is used to ionize hydrogen and helium during the contraction. The amount of released recombination energy during each pulsation cycle increases exponentially because the locations of the partial ionization zones, more specifically their mass coordinates, vary with an exponentially growing amplitude. The total released recombination energy per cycle is directly proportional to the difference in the mass coordinate of the partial ionization zones at maximum compression compared to maximum expansion. At $t = \qty{32.5}{yr}$, the envelope extends to such large radii that about \qty{2}{\msun} of the envelope are completely neutral. Additionally, the recombination front of doubly ionized helium is about \qty{4}{\msun} below the surface of the star. At this time, the dynamical timescale of the envelope, approximated by 
\begin{equation}\label{eq:tau_dyn}
    \tau_\mathrm{dyn} = \sqrt{\frac{R^3}{G M_\mathrm{env}}},
\end{equation}
is longer than the radiative cooling timescale of the envelope (Fig.~\ref{fig:restructuring}d), approximated by
\begin{equation}\label{eq:tau_KH}
    \tau_\mathrm{KH} = \frac{G M M_\mathrm{env}}{2 R L}.
\end{equation}
Here, $R$ is the radius of the RSG, $M$ is its mass, $G$ is the gravitational constant, $M_\mathrm{env}$ is the mass of the convective envelope, and $L$ is the luminosity of the RSG. Whenever $\tau_\mathrm{KH} < \tau_\mathrm{dyn}$, the envelope is losing energy via radiation faster than it can respond dynamically \citep[c.f.][]{clayton2017a}. This causes a catastrophic cooling event at the surface, where a significant fraction of the internal energy of the envelope is radiated away through the optically-thin outer \qty{2}{\msun} layers. As a consequence, the total energy (sum of potential, internal, and kinetic energy) of the convective envelope decreases (Fig.~\ref{fig:restructuring}e).

In the outer \qty{2}{\msun} layers, the opacity is orders of magnitude lower compared to the rest of the envelope, because hydrogen and helium are fully recombined ($\log(\kappa / \mathrm{cm^2 g^{-1}}) = 1.8$ in the hydrogen partial-ionization zone versus $\log(\kappa / \mathrm{cm^2 g^{-1}}) < -3$ in the neutral layers). Therefore, the neutral layers are stable against convection. As the expanded envelope contracts, the neutral, radiative layers become re-ionized and get mixed with the convective layers further below. This mixes the low-entropy radiative layers with the high-entropy convective layers further inside, lowering the average entropy in the entire convective envelope. Additionally, we observe that the entropy gradient changes during the restructuring event. The entropy gradient becomes less negative, which can be seen by a smaller spacing between the lines of constant mass after the restructuring in Fig.~\ref{fig:restructuring_entropy}. As a consequence of the altered thermal structure of the envelope, the density structure adjusts accordingly. We find on average larger densities deep in the convective envelope and lower densities closer to the surface compared to the original hydrostatic structure. The shallower entropy gradient might also have consequences for the efficiency of the convective energy transport. Such restructuring phases have been previously found and are described in \citet{yaari1996a}, \citet{lebzelter2005a}, and \citet{clayton2018a}. They seem to be unavoidable for pulsating envelopes once the dynamical and thermal timescales become comparable, even if the initial growth of the pulsations is hindered (see Appendix~\ref{app:damping}). 

We expect the catastrophic cooling events would be observable as an infrared transient because of the large amount of energy radiated away in a short duration, and because of the cool effective temperature of the RSGs. For the \qty{15}{\msun} RSG model, the maximum luminosity during the catastrophic cooling event reaches \qty{7\times 10^{40}}{erg\,s^{-1}}. In total, $5\times 10^{46}$ ($7\times 10^{46}$)~erg are radiated away during a period of $30$ ($120$)~days. Such timescales and energetics are very similar to observed luminous red novae \citep[\eg][]{pastorello2021a}.

After the restructuring phase, the pulsations continue at smaller but varying amplitude ($\qty{35}{yr} \lesssim t \lesssim \qty{80}{yr}$). During this phase, the envelope is thermally and dynamically adjusting to the structural changes of the catastrophic cooling event. At a time of around \qty{80}{yr}, the density and entropy converge to a new structure, indicating that the envelope has reached a new thermal and dynamical equilibrium state.

Once the envelope is adjusted to the changes after about \qty{80}{yr}, the pulsations become periodic. This phase lasts for several hundred years. Because the initial \qty{80}{yr} of the hydrodynamic simulation is still affected by the switching from the hydrostatic to the hydrodynamic model, it seems unlikely that this phase represents the behavior of the true stellar envelope. In a more realistic scenario, we envision that the envelope starts to pulsate gradually once the star climbs the RSG branch. Thus, it increases its $L/M$ ratio and experiences the envelope restructuring once the amplitudes are large enough for the catastrophic cooling event to set in. The catastrophic cooling event itself is unavoidable because with the growing amplitude of the pulsations, the ratio of the thermal to dynamical timescale decreases, eventually becoming on the order of unity, and triggering a thermal restructuring of the envelope. The exact path of reaching the catastrophic cooling event is expected to occur over a much longer timescale compared to the one we simulate here. We find that the steady-state pulsations afterwards are identical, even if the evolution and the growth of the pulsations that lead to the catastrophic cooling are changed, for example by applying a damping force in the envelope (see Appendix~\ref{app:damping}). Hence, we focus on the steady-state pulsations with $t > \qty{80}{yr}$ in the following analysis of the pulsations.

\subsection{Periodic pulsations of RSG envelope}\label{sec:pulsations:mechanism}

\begin{figure*}
    \centering
    \includegraphics[width=17cm]{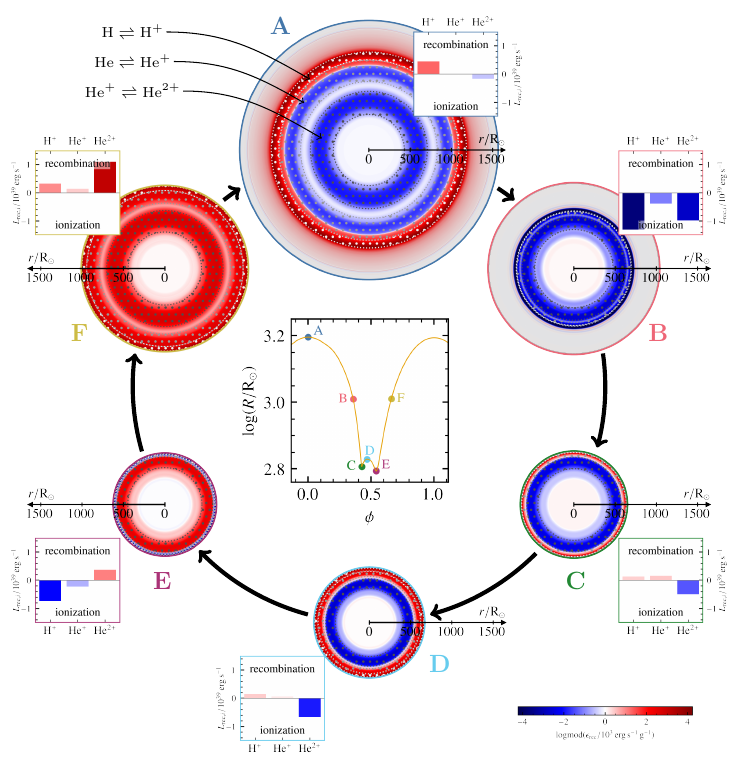}
    \caption{Recombination structure of the RSG envelope at points A--F. The six colored circles show the radius of the RSG at points A--F. The coloring inside the circles shows the specific released recombination energy $\epsilon_\mathrm{rec}$ as $\mathrm{logmod}(\epsilon_\mathrm{rec} / 10^3\,\mathrm{erg\,s^{-1}\,g^{-1}})$ on a linear radial scale, with $\mathrm{logmod}(x) = \mathrm{sign}(x)\log_{10}(|x|+1)$ \citep{john1980a}. Red corresponds to energy released by recombination, while blue corresponds to energy being removed via ionization. The square insets at each point show the recombination luminosity $L_{\mathrm{rec},i} = \int_{M_\mathrm{env}} \epsilon_{\mathrm{rec},i} \,\mathrm{d}m$ for the species $i = \mathrm{H^+,\ He^+}$ and He$^{2+}$ on a linear scale. Hatched regions indicate the partial ionization zone for hydrogen and helium, defined where the ionization fraction of the corresponding species is between \qty{1}{\%} and \qty{99}{\%}. Light-gray shading at points A and B corresponds to neutral layers (ionization fractions less than \qty{1}{\%}). The central plot shows the radius evolution of one pulsation cycle and indicates the 6 highlighted points A--F. The temporal evolution of the recombination structure is available as an \href{https://zenodo.org/records/16876815/preview/Fig4_recombination_evolution.mp4?include_deleted=0}{online movie}.}
    \label{fig:pulsations_cartoon}
\end{figure*}

For $t > \qty{80}{yr}$, the radial pulsations are periodic with a pulsation period of $817^{+8}_{-11}\,\mathrm{d}$ that changes by less than $\qty{2}{\%}$ during the simulated time. This period is determined using a Lomb-Scargle periodogram and the uncertainty is the full width at half maximum. For comparison, the initial dynamical timescale is about \qty{230}{d}. A zoom on one pulsation cycle at \qty{102}{yr}--\qty{104}{yr} is shown in Figs.~\ref{fig:pulsations_HRD}b and~\ref{fig:pulsations_HRD}c. The luminosity varies by more than one order of magnitude during the pulsation cycle ($4.45 \leq \log(L/\lsun) \leq 5.90$) and the effective temperature varies between $2200$ and \qty{6820}{K} (\ie, $\log(T_\mathrm{eff}/\mathrm{K}) = 3.342$ and $3.834$, respectively). These variations are much larger than typical uncertainties on supernova progenitor observations ($\Delta \log(L/\lsun) \approx 0.2$ and $\Delta \log(T_\mathrm{eff}/\mathrm{K})\approx 0.05$; \citealt{smartt2015a}). 

We indicate six points during the pulsation cycle (A--F), which are chosen to highlight the characteristic stages during one pulsation cycle. Point A is chosen to be at the maximum radial extent. Points B and F are chosen to be close to the hydrostatic radius of the RSG at $1024\, \rsun$, with point B in the contraction phase and point F in the expansion phase of the pulsation. Points C and E are chosen to be at the local minima of the radial extent of the RSG. These two points are separated by a small expansion phase that lasts less than \qty{100}{d}. Point D is chosen to be at the local maximum radius between points C and E.

The pulsations are mainly driven by the combination of the $\kappa$- and $\gamma$-mechanism ($\kappa\gamma$-mechanism). The former is connected to the changing opacity in partial ionization layers of hydrogen and helium, and the latter corresponds to the released/consumed energy by a change in the ionization state of hydrogen and helium during the expansion/contraction. The entire driving mechanism is visualized in Fig.~\ref{fig:pulsations_cartoon}, which shows the recombination structure of the RSG at points A--F.

At point B, the entire envelope is in a contraction phase and both hydrogen and helium are being re-ionized (blue color in Fig.~\ref{fig:pulsations_cartoon}). The process of re-ionization takes energy from the pulsation and converts it to ionization energy ($\gamma$-mechanism). Additionally, the partial ionization zones of hydrogen and helium increase the opacity, therefore increasing the radiation pressure/Eddington factor, which opposes the contraction of the envelope ($\kappa$-mechanism; see Appendix~\ref{app:pulsations}).

As the envelope contracts, hydrogen and helium get completely re-ionized. At point C, the entire envelope is ionized. The outermost layers fall back onto layers of larger density and pressure; they get compressed, bounce off, and re-expand (\cf~Figs.~\ref{fig:pulsations_kipp1} and \ref{fig:pulsations_kipp2}). As they cool down during re-expansion, H$^+$ and He$^+$ start recombining (red color in Fig.~\ref{fig:pulsations_cartoon}) and drive the expansion of the surface layers. Only the outer \qty{0.1}{\msun} are expanding, while the layers further inside the envelope keep on contracting and cause further re-ionization of He$^{2+}$. Because only the immediate surface layers are expanding, the expansion phase starting at point C reaches only a small maximal radial extent at point D of \qty{675}{\rsun}. At this point, all the recombination energy from H and He$+$ of the expanding layers is released. Another contraction phase follows, during which the surface layers are again re-ionized. Meanwhile, layers further inside the envelope keep contracting and re-ionizing He$^{2+}$ until enough pressure (both gas and radiation pressure) builds up to stop the contraction and start the expansion of the interior envelope layers.

At point E, the contracting surface collides with the expanding interior envelope layers, reaching another minimum in total radial extent. This point marks the smallest radial extent throughout the entire pulsation cycle with $R = \qty{622}{\rsun}$. The interior layers are accelerated by the recombination of He$^{2+}$, which marks the main difference to point C, where He$^{2+}$ is being re-ionized. 

At point F, the whole envelope expands. The envelope cools down, which causes the recombination of hydrogen and helium. The released recombination energy further accelerates the expansion, driving it to large radii. Most of the released recombination energy is from He$^{2+}$ with smaller contributions of hydrogen and He$^{+}$. The recombination of helium stops already before reaching the maximum extent of the envelope at point A, while the recombination of hydrogen continues for a longer time, giving the surface layers an extra push.

At point A, the star is at its maximal radial extent of \qty{1568}{\rsun}. Now, the surface layer starts to contract. However, because of hydrogen recombination, there is a subsurface layer with an outward radial velocity causing a density inversion (Fig.~\ref{fig:density_profiles}). Layers below the hydrogen recombination zone are contracting and heating up, leading to the re-ionization of helium. As a result, the entire envelope contracts, and we are again at the state in point B. In this manner, the pulsation cycle driven by this $\kappa\gamma$-mechanism is sustained and remains stable.

\subsection{Density variations by the pulsations}\label{sec:pulsations:density}

\begin{figure*}
    \centering
    \includegraphics[width=17cm]{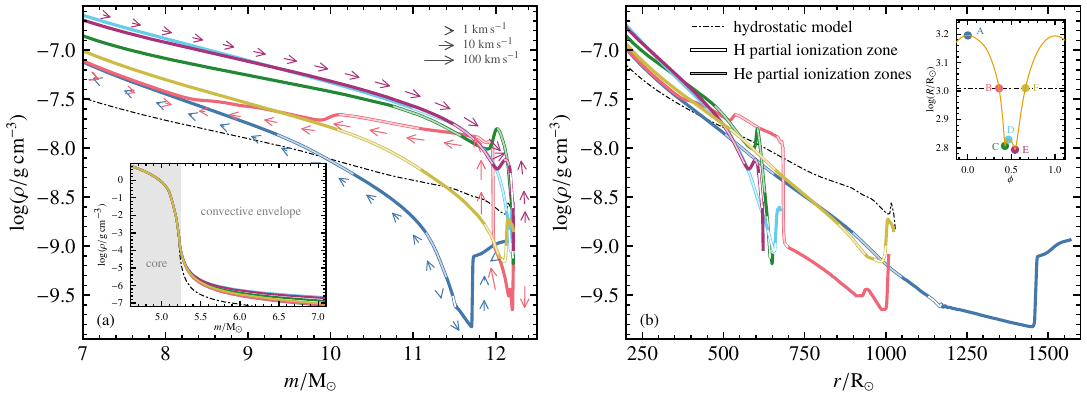}
    \caption{Density profiles of the RSG envelope at points A--F during the pulsation cycle in terms of mass coordinate (panel a) and radius coordinate (panel b). The highlighted regions show hydrogen and helium partial ionization zones where the ionization fraction is between \qty{10}{\%} and \qty{90}{\%}. The density profile of the hydrostatic model is shown for comparison. Arrows along the profiles indicate the radial velocity, both the direction (negative/inwards or positive/outwards) and the magnitude, and highlight layers that are expanding/contracting. The inset in panel (a) shows the density structure of the interior layers, up to where the core was cut out. The shading indicates the core region of the RSG. The inset in panel (b) shows the radius evolution of one pulsation cycle and indicates the points A--F. The temporal evolutions are available as an \href{https://zenodo.org/records/16876815/preview/Fig5a_density_profile_evolution_mass.mp4?include_deleted=0}{online movie panel a} and an \href{https://zenodo.org/records/16876815/preview/Fig5b_density_profile_evolution_radius.mp4?include_deleted=0}{online movie panel b}.}
    \label{fig:density_profiles}
\end{figure*}

The large amplitude pulsation of the RSG, driven by the $\kappa\gamma$-mechanism, significantly alters the density structure of the envelope (Fig.~\ref{fig:density_profiles}). For mass coordinates below \qty{10}{\msun}, we find consistently larger densities compared to the hydrostatic model. This is a consequence of the catastrophic cooling event at $t=\qty{32.5}{yr}$ that reduces the total energy of the envelope and leads to the rearrangement of the envelope structure. As expected, we find lower densities for the models in the expanded states compared to the models in the compressed states. The largest densities in the deep interior are reached at point D. Although the stellar radius of point D is larger than at points C and E, the deeper layers keep contracting between points C and D, causing the density to increase.

The density near the surface ($m>\qty{11.5}{\msun}$) shows a lot of structure, and we find large jumps in density. Most of this structure can be directly connected to the recombination/re-ionization processes happening in the envelope. There is a large density inversion at $m = \qty{11.7}{\msun}$ ($r = \qty{1450}{\rsun}$) in the density profile of point A. Hydrogen recombination at \qty{11.5}{\msun} accelerates low-density material that crashes into high-density surface layers, which are already contracting. The high-density surface layer is built up by the continuous expansion driven by hydrogen recombination. At point B, there is a large drop in density at \qty{12}{\msun}, which is an immediate consequence of the surface layers falling back onto the deeper, ionized layers. A shock front develops at the interface where the neutral surface layers are re-ionized. The post-shock density is about ten times larger than the pre-shock density (\qty{10^{-8}}{g\,cm^{-3}} post-shock versus \qty{10^{-9}}{g \, cm^{-3}} pre-shock). At points C and E, we find smaller density inversions just below the surface. Layers above the density inversion participate in the short-lived expansion phase between points C and E, while layers further below keep contracting and increasing their density. At point F, we find a smooth density profile, as the whole envelope is expanding coherently. However, just below the surface, there is already a pile-up of high-density material from hydrogen recombination. This layer keeps on growing further until point A. Note that the density profiles at points F and B show very different structures, even though the radius of the star is the same.

In the layers below the convective envelope, the density structure is unchanged because the pulsations are confined to the envelope. We find variations of less than \qty{0.5}{\%}, which decrease deeper inside the star, indicating that the pulsations are damped in these layers.

\subsection{Dust models of RSG pulsations}\label{sec:pulsations:dusty}

\begin{figure}
    \centering
    \resizebox{\hsize}{!}{\includegraphics{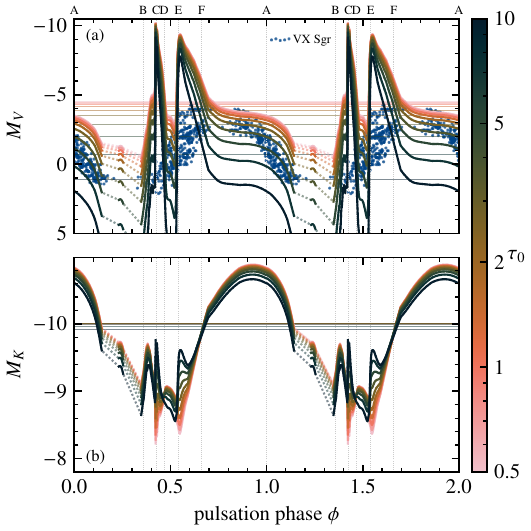}}
    \caption{$V$- and $K$-band light curves of the pulsating RSG surrounded by a dust shell. The parameter $\tau_0$ sets the optical depth at the beginning of the pulsation cycle ($\phi = 0$). The light dotted lines indicate regions outside the MARCS grid (too low $T_\mathrm{eff}$) that do not allow us to make photometric estimates. For better visualization, two full pulsation cycles are shown. The light, horizontal lines show the time-averaged magnitudes. For comparison, the phase-folded, visual light curve of VX~Sgr is shown as well, with data taken between JD=2640000 and 2644000, and assuming a period of $747\, \mathrm{days}$.}
    \label{fig:dusty_LC}
\end{figure}

We use the DM models (see Sect.~\ref{sec:methods:dust}) to construct light curves of the pulsating RSG with varying amounts of dust around it. The light curves in the $V$ and $K$ bands are shown in Fig.~\ref{fig:dusty_LC}. 

The light curves do not follow a sinusoidal modulation of the brightness, but show a rather complex morphology. We find two short-duration peaks, corresponding to points C and E (\ie, the states of maximum compression). Around the maximum extent of the RSG (point A, $\phi=0$), the light curves vary more smoothly. In the $V$-band light curves, we find that the brightness is slowly decreasing and might even reach a plateau phase, depending on the amount of dust. In the $K$-band light curves, this phase corresponds to the maximum light of the star. Unfortunately, we are unable to predict the cool phases of the light curves because the underlying MARCS models are only available for $T_\mathrm{eff} \geq \qty{2400}{K}$.

The amount of dust, modeled via different optical depths $\tau_0$, has a significant influence on the amplitude of the $V$-band light curve. We vary $\tau_0$ between $0.5$ and $10$, which corresponds to total dust masses between $10^{-4}$ and $10^{-2}\,\msun$, inline with expectations of dusty RSGs such as VY~CMa \citep{humphreys2022a}, NML~Cyg \citep{debeck2025a}, or the progenitor of SN~2023ixf \citep{niu2023a}. For low amounts of dust, we find a total brightness variation of $\approx \qty{10}{mag}$, while for increasing amounts of dust, we find a brightness variation of up to \qty{25}{mag}. The $K$-band light curves are less influenced by the amount of dust around the RSG. The brightness variation for the $K$ band is around 2.0 to \qty{2.5}{mag} with increasing amounts of dust. 

We find that for the $V$ band, the more dust is present around the RSG, the lower is the brightness at any time during the pulsation. This trend does not hold for the $K$-band light curve. While the average brightness decreases with increasing amounts of dust, we also find times during the pulsation cycle (\eg, point E at $\phi=0.55$) where the opposite ordering is the case. That is, we find that the RSG is brighter in the model with more dust. 

For comparison, we show the visual light curve of VX~Sgr\footnote{Taken from the American Association of Variable Star Observers (AAVSO) database at \url{https://www.aavso.org/data-download}, with credits to the Royal Astronomical Society of New Zealand (RASNZ).}, a semi-regular pulsating RSG with a large amplitude in the optical. The peak-to-peak variability of VX~Sgr is about $6-7\,\mathrm{mag}$, which is lower than the dust free ($\tau_0 \rightarrow 0$) peak-to-peak variability of about $10\,\mathrm{mag}$ of the model. Note that the observations of VX~Sgr are visual magnitudes, and the computed light curves are in the $V$-band. There is also some despute whether VX~Sgr is a RSG, an extreme AGB star or even a Thorne-\.{Z}ytkov object \citep{tabernero2021a}. In any case, the envelope structure should be very similar and comparable to the $\qty{15}{\msun}$ RSG model. For a more detailed comparison to observations, see Sect.~\ref{sec:discussion:obs}.

\section{Supernova light curves}\label{sec:LC}

We model the supernova light curves for the RSG at various stages throughout the pulsation cycle. Although the pulsation simulations are computed at the end of core carbon burning, we expect the pulsations to continue until the onset of core collapse because the dynamical and thermal timescales of the envelope are larger than the nuclear timescale of the core. Therefore, we do not expect that the timing of the core collapse is in any way related to the pulsation phase of the envelope. This means core collapse can happen at any time during the pulsation cycle of the envelope with equal probability. We highlight the supernova light curves corresponding to explosions at points A--F as representative states of the envelope during one pulsation cycle.

\subsection{Light-curves at different pulsation phases}

\begin{figure}
    \centering
    \resizebox{\hsize}{!}{\includegraphics{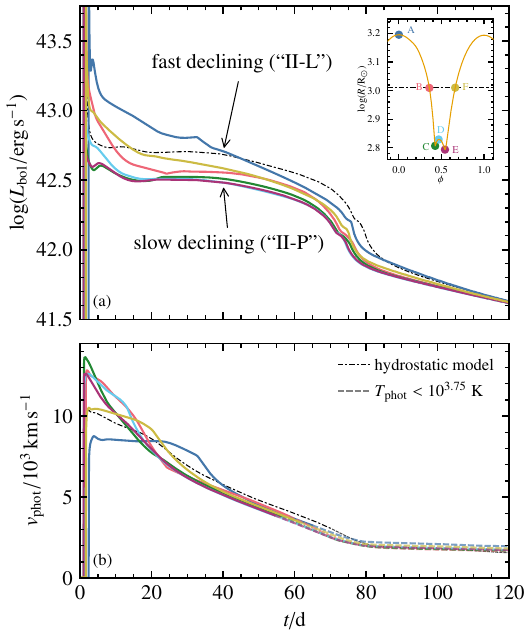}}
    \caption{Bolometric light curves (panel a) and photospheric velocities (panel b) for supernovae at points A--F during the pulsation cycle. The explosion of the hydrostatic model is shown for comparison. The inset shows the radius evolution of one pulsation cycle and indicates the points A--F. Once the photospheric temperature drops below $10^{3.75}\,\mathrm{K}$, the velocities are no longer reliable because \texttt{SNEC} cannot accurately estimate the location of the photosphere due to limitations in low-temperature opacities \citep{morozova2015a}. The complete $\phi$-evolution of the light curve and the photometric velocity is available as an \href{https://zenodo.org/records/16876815/preview/Fig7_light_curve_evolution.mp4?include_deleted=0}{online movie}.}
    \label{fig:LC_vphot}
\end{figure}

We simulate the supernova explosion of the red supergiant at different phases of the pulsation with the same explosion parameters, derived by applying the model of \citet{muller2016a} (see Sec.~\ref{sec:methods:LC}). The resulting light curves for each pulsation phase are shown in Fig.~\ref{fig:LC_vphot}a. We find that the light curve shapes and decline rates are strongly impacted by the pulsation of the red supergiant. At all phases of the pulsation, the obtained light curves are different from the typical Type~II-P shaped light curve obtained by simulating the explosion of a hydrostatic red supergiant model, which is the standard procedure in this field (see the dash-dotted line in Fig.~\ref{fig:LC_vphot}a). At large radii (point A), the light curve is declining fast during the first \qty{80}{d} ($1.1\,\mathrm{dex} / 100\,\mathrm{d}$ in $L_\mathrm{bol}$ or $3\,\mathrm{mag} / 100\,\mathrm{d}$ in the $V$~band\footnote{\texttt{SNEC} computes absolute magnitudes in different filters assuming blackbody emission from the photosphere and bolometric corrections from \citet{ofek2014b}.}), and might have classically been referred to as a Type~II-L SN. At later times, the light curve shows the characteristic exponential decay when the radiation output is dominated by the contribution of radioactive $^{56}\mathrm{Ni}$. In contrast, at small radii (points C--E) the light curve has the long ($\sim 80$ d) plateau phase, \ie small decline rates ($0.45\,\mathrm{dex} / 100\,\mathrm{d}$ in $L_\mathrm{bol}$ or $1.37\,\mathrm{mag} / 100\,\mathrm{d}$ in the $V$~band), characteristic of II-P supernovae, and reaches lower luminosities. At intermediate radii (points B and F), when the star has approximately the hydrostatic radius, the light curve shows an early excess emission followed by a decline with decline rates between the two extremes shown above.

We show the corresponding photospheric velocities $v_\mathrm{phot}$ in Fig.~\ref{fig:LC_vphot}b. There is a clear distinction between points A and F, and all other points. For both points A and F, we find that after the shock passed through the ejecta, $v_\mathrm{phot}$ stays constant for up to one month before decreasing. At the other points, $v_\mathrm{phot}$ peaks directly after shock breakout and then steadily declines. At these points, the maximum $v_\mathrm{phot}$ is also the highest. 

Both effects, the higher luminosities and the lower $v_\mathrm{phot}$ at points A and F compared to the other points, can be directly connected to the different envelope structures at the time of explosion. At points A and F, the envelope is extended to beyond $1000\,\rsun$. This leads to more deceleration compared to the compact progenitors, \ie, lower photospheric velocities. The higher the deceleration, the more explosion energy is converted to radiation, which leads to a higher luminosity. At point B, the radius of the progenitor is the same as point F, but the density structure is different, having larger densities in the outer layers of the envelope (Fig.~\ref{fig:density_profiles}). This results in a different light curve and photosperic velocity at point B compared to point F.

In summary, we find that taking the hydrodynamical changes of red supergiants into account, \ie, their radial pulsations driven by the $\kappa\gamma$-mechanism, naturally reproduces some of the diversity of hydrogen-rich supernova light curves \citep{anderson2014a}

\subsection{Connecting light curve features to the progenitor structure}

\begin{figure*}
    \centering
    \includegraphics[width=17cm]{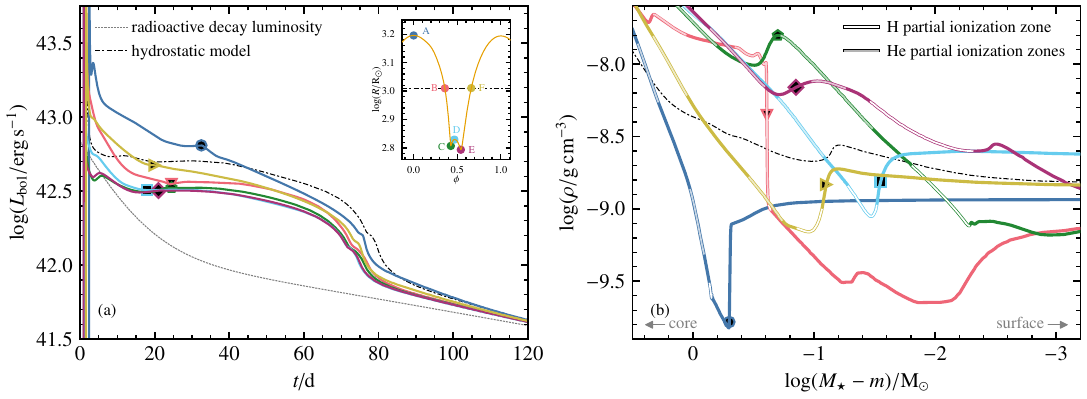}
    \caption{Connecting feature in the supernova light curves to features in the density profiles of the RSG envelope. Panel (a) is similar to Fig.~\ref{fig:LC_vphot}a and panel (b) is similar to Fig.~\ref{fig:density_profiles}. The density profiles are shown as a function of the mass until the surface. Colored markers connect features in the density profiles to features in the supernova light curves. Features in the light curve arise when the SN photosphere reaches the mass coordinates highlighted in the density profile.}
    \label{fig:LC_profiles}
\end{figure*}

As noted in the previous section, the light curves we obtain at different pulsation phases have different shapes. They also display features, such as the short plateau-like feature of about \qty{10}{d} at point A. We investigate the origin of these shapes and features by studying the connection to the progenitor structure in Fig.~\ref{fig:LC_profiles}; in panel b of this figure, we present the pre-supernova density profiles at different phases of the evolution. These are highly affected by the ionization and recombination states described in Sec.~\ref{sec:pulsations:density}. As expected, at the point of the largest radius expansion (point A), the stellar structure has a particularly low density in the envelope, while at the point when the smallest radius is reached, it has the highest overall density. The density profiles show prominent density inversions near the surface that are directly related to the ionization structure (see Sec.~\ref{sec:pulsations:density}).
 
While the supernova shock goes through the stellar structure, it heats and accelerates the layers it encounters. Once it has traversed most of the entire stellar structure and the first photons escape, the supernova light curve begins. For different pulsation phases, the progenitor size varies greatly. This is reflected by the time it takes until the peak luminosity is reached, which increases for larger RSG radii (light curves A vs. C in Fig.~\ref{fig:LC_profiles}).

The supernova shock leaves the ejecta completely ionized, and the opacity increases greatly in the layers that were originally recombined. Photons are trapped behind the optically-thick layers. As a cooling mechanism, the ejecta expands rapidly and the temperature and luminosity drop, leading to a drop in the light curves. At this point, the photosphere remains at the mass coordinate of the surface of the pre-explosion star (\cf~Fig.~\ref{fig:phot_props}). The rate at which the luminosity drops reflects the state of the progenitor. The more expanded the outer layers, the lower the density, and the slower the initial decline rate of the light curve.

Once the hydrogen-rich ejecta have expanded and cooled sufficiently for the recombination of hydrogen atoms to occur, the photosphere moves inward, and the photons trapped behind it are released, contributing to the supernova light curve. In addition, the temperature decreases less rapidly. By following the location of the photosphere during the explosion, the features in the supernova light curves can be traced back to the moments when the photosphere reaches certain parts of the stellar layers. We highlight these moments by markers on the density profiles and supernova light curves in Fig.~\ref{fig:LC_profiles}. 

We find that when the photosphere reaches the layers where the density inversions connected to the partial-ionization zones occur, features in the light curves arise. At point A, there is a density inversion in the pre-SN density profile at $\log(M_\star - m)/\msun = -0.3$. After the SN, the photosphere reaches this particular mass coordinate at a time of \qty{30}{d}. At this particular time, we see a short plateau feature in the corresponding light curve (Fig.~\ref{fig:LC_profiles}). The density inversion at point A arises from the released recombination energy of hydrogen (see Sect.~\ref{sec:pulsations:density}). Similarly, at point B we find that the density increases rapidly by one order of magnitude at $\log(M_\star - m)/\msun = -0.6$ because of the re-ionization of hydrogen and helium. This particular mass coordinate is reached by the photosphere \qty{25}{d} after the explosion and marks the transition of the light curve from the near-linear decline to the plateau phase. A similar feature is visible in the light curves corresponding to point F, where the decline rate changes about \qty{20}{d} after the explosion. This correlates with the density inversion found at $\log(M_\star - m)/\msun = -1.1$, caused by hydrogen recombination. Point D behaves qualitatively similar to point F. The density inversion at $\log(M_\star - m)/\msun = -1.5$ is caused by hydrogen recombination during the intermediate expansion phase. When the photosphere reaches this mass coordinate at about \qty{18}{d} after the explosion, the decline rate changes (\ie, from linear decline to plateau). Points C and E show only weak density inversions that get washed out by the supernova shock. When the photosphere passes these mass coordinates, no significant changes in the light curves appear. 

\section{Discussion}\label{sec:discussion}

\subsection{When do the pulsations occur?}\label{sec:discussion:when}
It has been shown by \citet{heger1997a}, \citet{yoon2010a}, and also \citet{clayton2018a} that the growth rate of the pulsations increases with increasing luminosity-to-mass ratio $L/M$, which is closely connected to the Eddington factor. Therefore, we expect stronger pulsations as stars evolve in the RSG phase because they typically get brighter while losing mass via winds. For the \qty{15}{\msun} RSG model that we analyzed, we did not find any pulsations during core helium burning. The pulsations only become significant once the star evolves to burn carbon in its center. Based on the work by \citet{heger1997a} and \citet{yoon2010a}, weaker pulsations are expected in lower mass RSGs to occur only shortly before core collapse. For higher mass RSGs, the pulsations are expected to start earlier in the evolution and become more extreme. 

\citet{clayton2017a} showed that similar pulsations might also be present in the context of common-envelope evolution. They find that depending on the energetics of the common-envelope event, the pulsations might even lead to the dynamical ejection of the outer envelope.

The pulsation mechanism itself depends only weakly on the metallicity of the star, because the partial ionization zones of hydrogen and helium are mainly responsible for driving the pulsation. Therefore, RSGs of similar mass as considered here but with a different metallicity should behave similarly regarding the pulsations. However, the mass loss rate during earlier evolutionary phases might heavily depend on the metallicity \citep[\eg][]{puls2008a,mokiem2007a}. In such cases, the mass loss has typically only a minor effect on the luminosity of the star. Hence, the $L/M$ ratio of higher metallicity stars is larger, and we expect even stronger pulsations.

For the \qty{15}{\msun} RSG, a supernova with a fast-declining light curve (classically Type~II-L) is more likely compared to a slow-declining (Type~II-P) light curve, because the RSG spends more time in the extended phase. Observationally, many more II-P SNe are found compared to fast-declining (II-L) SNe amongst all core-collapse SNe -- \citet{smartt2009a} report \qty{59}{\%} II-P and \qty{3}{\%} II-L while \citet{smith2011a} report \qty{48.2}{\%} II-P and \qty{6.4}{\%} II-L. Therefore, it might seem as if our model over-predicts the numbers of fast-declining SNe. However, because of the initial-mass function \citep[e.g.][]{salpeter1955a,kroupa2001a}, lower-mass RSGs are more common SN progenitors compared to higher-mass RSGs. As described above, we do not expect lower-mass RSGs to pulsate significantly at core collapse \citepalias{paperII}. Hence, we would expect most lower-mass RSGs to explode as Type~II-P SNe only. Extending on this, we predict that the progenitors of fast-declining (Type~II-L) SNe, that originate from this mechanism, are all massive RSGs.

\subsection{Mass loss during the RSG phase}
Inferring the winds of RSGs either empirically or via models is challenging. This has led to a plethora of RSG wind prescriptions (\citealt{dejager1988a,nieuwenhuijzen1990a,vanloon2005a,beasor2020a,kee2021a,antoniadis2024a}; for a review, see \citealt{smith2014a}). These prescriptions can differ by more than one order of magnitude. In our models, we assume that the RSG wind mass loss follows the prescriptions of \citet{dejager1988a}. Varying the wind prescription in the RSG phase might have dramatic effects on the pulsations of the envelope. For higher/lower winds, the total convective envelope mass before the onset of core collapse can be lower/higher. This can impact the density structure of the envelope and $L/M$, which in the end are responsible for the exact morphology of the pulsation and possibly also the pulsation period \citep{gough1965a,heger1997a}. 

It has also been suggested that the pulsations themselves are responsible for the high mass loss rates of RSGs. \citet{yoon2010a} suggest that the pulsations can drive a superwind during the final evolution as an RSG. These winds would increase as the pulsations become stronger, \ie, in more massive and more evolved RSGs. Alternatively, dynamical mass ejections remove some envelope material on a short timescale \citep{tuchman1978a,clayton2017a,clayton2018a}. Observationally, the highest mass loss rates in RSGs are found for objects like VY~CMa \citep{humphreys2022a}, VX~Sgr \citep{decin2024a}, and NML~Cyg \citep{debeck2025a}, which all show significant pulsations \citep[\eg][]{monnier1997a,kiss2006a}. 

In either case, the enhanced mass loss leads to a larger $L/M$ ratio, because the luminosity at the RSG stage is only determined by the carbon-oxygen core mass \citep[\eg][]{temaj2024a}. This leads to more extreme pulsations that continue to remove the envelope. Depending on the overall strength of such winds or mass loss, the RSG might be able to evolve again towards hotter temperatures once the mass of the hydrogen-rich envelope decreases below some critical value. Then, the envelope is no longer unstable to pulsations, and the star might end its life as a yellow supergiant. In this case, an explosion as a Type~IIb or even Type~Ib is expected.

For simulations of the $15\,\msun$ RSG presented above, we turn off any mass loss during the dynamical evolution. Because the luminosity and radius of the star change by up to one order of magnitude during one pulsation cycle, any wind determined by classical wind mass-loss prescriptions \citep[\eg,][]{dejager1988a} would be modulated directly by the changes of these surface properties. However, it is unclear whether such prescriptions provide a meaningful wind mass-loss for dynamical models. Therefore, we cannot make statements about pulsationally-driven superwinds, but refer to the results obtained by \citet{yoon2010a}. We do not find any dynamical mass ejection for the $15\,\msun$ simulations, but expect these to set in for more massive RSG with larger $L/M$. The overall effect of successive mass ejections might then be similar to a pulsationally-driven superwind, producing a shell-like structure of CSM around the RSG, similar to the CSM observed around VY~CMa \citep{singh2023a} and NML~Cyg \citep{andrews2022a}, or thermally-pulsing asymptotic giant branch stars \citep[\eg][]{olofsson1988a,mattsson2007a}. Additionally, such mass loss should only weakly depend on the metallicity of the star, because the pulsation mechanism described in Sect.~\ref{sec:pulsations:mechanism} does not rely on the metal content in the envelope of the RSG.

In any case, all these effects might leave some CSM around the SN progenitor. This is true especially for more massive RSGs, as they exhibit the strongest pulsation, in which we expect dynamical mass ejections \citep{clayton2018a}. This has direct implications for the missing red supergiant problem \citep{smartt2009b,smartt2015a} and could explain the maximum luminosity observed in RSGs \citep{davies2018a}.

The supernova ejecta is also expected to interact with the CSM that is produced by enhanced mass loss during the pulsation. The CSM interaction typically leads to excess emission during the early light curve \citep{moriya2011a}, as observed in many Type~II SNe \citep{anderson2024a}. Depending on the exact nature of the CSM, the star might even produce Type~IIn SNe, \ie, hydrogen-rich supernovae with narrow line emission features in the spectra. Including the CSM effects on the light curves is beyond the scope of this paper, but is discussed in detail in \citetalias{paperII}.

\subsection{Pulsating RSGs in binary stars}
If a pulsating RSG is part of a binary star system, many evolutionary scenarios can open up, depending on the initial separation \citep[\eg][]{podsiadlowski1992a,podsiadlowski1993a,nomoto1993a,nomoto1995a,ercolino2024a,dessart2024b}. There could be episodic mass transfer to the companion star, possibly leading to a complex and shell-like CSM structure \citep{landri2024a}. The evolution of the binary orbit is most likely very sensitive to the initial conditions of such an episodic mass transfer phase, and it is plausible that such a phase might increase/decrease the eccentricity of the binary depending on how the pulsation period relates to the orbital period and how the pulsation period changes upon mass transfer. Naively, one expects the pulsations and orbital timescale to be similar upon mass transfer, as the pulsations are a dynamical phenomenon and the dynamical timescale is similar to the orbital period.\footnote{\citet{gough1965a} has shown that for a fully convective star, the pulsation period $P$ follows $P\propto R^2/M$ rather than the dynamical timescale (see Eq.~\ref{eq:tau_dyn}).} However, the interaction between mass transfer and the pulsations may result in a non-trivial evolution of the system.

\subsection{Comparison to observed pulsating RSGs}\label{sec:discussion:obs}
There is a large number of observed RSGs that show period variability \citep{kiss2006a}. Based on theoretical arguments, the pulsation period of RSGs is supposed to scale linearly with $L/M$ \citep{gough1965a}. From observations, period-luminosity (PL) relations can be measured to use RSGs as standard candles for distance estimation \citep[\eg][]{jurcevic2000a}. Practically, the absolute magnitude in the $K$ band, $M_{K}$, is used because it is more accessible than the bolometric luminosity, and it is only little affected by dust attenuation. Based on the models of the \qty{15}{\msun} RSG with no dust, we find a period of $817^{+8}_{-11}\,\mathrm{d}$ and $\langle M_{K}\rangle = \qty{-10}{mag}$ (Fig.~\ref{fig:dusty_LC}b). \citet{chatys2019a} used RSGs in the Galaxy and the Large Magellanic Cloud to derive a PL relation. For a period of \qty{817}{d}, the PL relation would expect $M_{K}$ to be between \qty{-9.5}{mag} and \qty{-11.5}{mag}. Using the PL relation from \citet{soraisam2018a} based on RSGs in M31, we would expect $M_{K} \approx \qty{-11.2}{mag}$ with a spread of about \qty{0.5}{mag}. Lastly, using the PL relations from \citet{ren2019a}, derived from RSGs in M33 and M31, one finds $M_{K} = -11.1 \pm 1.4 \, \mathrm{mag}$ (M33) and $M_{K} = -11.1 \pm \, 1.1 \mathrm{mag}$ (M31). These expectations from PL relations agree within the quoted uncertainties with our models.

Observed amplitudes of the $K$ band variability reach up to \qty{0.5}{mag} and tend to increase with pulsation period \citep{yang2018a,chatys2019a,soraisam2018a,ren2019a}. This trend is also suggested by theoretical models \citep{heger1997a,yoon2010a}. However, we do find larger amplitudes in our models, independent of the amount of dust around the RSG. The amplitudes we find in the $V$ band for models without any dust  are $\Delta M_V\approx\qty{5}{mag}$ and reach values of ${>}\, \qty{15}{mag}$ for considerable amounts of dust (high $\tau_0$). Some observed RSGs reach $\Delta M_V \approx 2-3\,\mathrm{mag}$ \citep[\eg S~Per, VY~CMa;][]{kiss2006a}, with VX~Sgr showing the largest brightness variations of up to $7\,\mathrm{mag}$. More typically, amplitudes of \qty{1}{mag} are found in the $V$ band \citep{levesque2007a}. There are multiple reasons for the discrepancy of the amplitude from our models with observation, such as the treatment of convection in 1D models (see~Sect.~\ref{sec:discussion:convection}). Additionally, our 1D models only allow for radial pulsations, but non-radial pulsations might be present in the envelopes of RSGs. These could lead to a decrease in the amplitude of the radial pulsations. Most importantly, the observed RSGs are all expected to be core-helium burning because of the much shorter timescale of any later burning phases. We only find these pulsations in the very late phases of evolution, \ie, core carbon burning and later. We expect that most of the observed RSGs in the Galaxy or nearby galaxies are currently in the core-helium burning phase because of the much longer timescale. The post-core carbon burning phase until core collapse lasts typically about $10^2\,\mathrm{yr}$, which is about $0.01\,\%$ of the entire RSG phase. Moreover, assuming that the pulsations only happen in the more massive RSGs ($> 12 \, \msun$), there is a chance of less than $1:10^4$ to catch a RSG in the phase where it is pulsating because of an unstable envelope. Because only a couple of $10^3$ RSGs have been monitored closely in the Milky Way and nearby galaxies \citep{dai2025a,dai2025b}, we expect that zero to one RSGs could have been caught with such large amplitude pulsations. Therefore, we do expect larger amplitudes in our models compared to the observations (see Sect.~\ref{sec:discussion:when}). We predict that if observed, such rare large-amplitude pulsations would be an indication of the imminent collapse of such stars. Massive RSGs that already pulsate with a large amplitude during the core-helium burning have likely already lost a significant amount of their envelope via dynamical mass ejections \citep{clayton2018a} or an enhanced wind \citep{yoon2010a} and no longer appear as an RSG, linking to the missing RSG problem.

Pre-explosion imaging of the progenitors to SN 2023ixf and SN 2024ggi show signs of periodical variability \citep[\eg,][]{jencson2023a,kilpatrick2023a,soraisam2023a,xiang2024a,xiang2024b}. In the case of SN 2023ixf, a dense and close circumstellar material can be inferred from early flash spectroscopy \citep[\eg][]{li2024a,zimmerman2024a}. Discussing the details of SN 2023ixf and SN 2024ggi within the context of our model of pulsations RSG and their impact on light curves is beyond the scope of this paper, but will be analyzed and discussed in detail in \citetalias{paperII}.

\subsection{Implications of envelope restructuring}
The envelope restructuring, observed at around $t=\qty{32.5}{yr}$, immediately changes the density structure of the envelope and deviates from the original, hydrostatic envelope structure by up to one order of magnitude. This also means that the pulsations before and after the restructuring event can change significantly, including the pulsation period \citep{yaari1996a,tuchman1978a}. Predicting post-restructuring pulsation properties from a hydrostatic structure before the restructuring based on linear analysis is impossible.

Usually, the hydrostatic stellar structure at the onset of core collapse is used in simulating supernova light curves. However, if such an envelope restructuring occurs before the onset of core collapse, hydrostatic models are unable to capture this effect, independent of how large the amplitude of the pulsations is.

\citet{goldberg2020a} modeled the fundamental and first overtone pulsation of an initially \qty{18}{\msun} RSG just before the onset of core collapse, and then studied their effect on the resulting supernova light curve. They find that the fundamental-mode pulsations do not significantly modify the SN light curve in unusual ways. The pulsations decrease (increase) the density in the entire envelope upon expansion (compression). These changes of the SN light curve from the fundamental-mode pulsations can be recovered with typical scaling relations \citep[\eg][]{popov1993a,goldberg2019a} by considering a progenitor with the same mass but larger (smaller) radius. They do report that the amplitude of the pulsations is still growing at the onset of core collapse. However, they do not find any restructuring of the envelope. This is most likely because they start simulating the pulsations at most \qty{1750}{d} (just over 3 full pulsation cycles) before the core collapses. This means that the pulsations did not have enough time to grow in amplitude such that a catastrophic cooling event can be triggered, which ultimately causes the envelope to restructure. The amplitudes of their pulsations are motivated by the observed brightness variations of RSGs. However, we stress that most observed pulsating RSGs are not beyond core-carbon burning (see Sect.~\ref{sec:discussion:when}), and, therefore, we do expect larger amplitudes at core collapse. 

In contrast to the fundamental-mode pulsations, \citet{goldberg2020a} find a significant deviation of the SN light curve compared to the hydrostatic progenitor when considering first-overtone pulsations. In this case, the density in the inner part of the envelope increases (decreases) while the density in the outer part of the envelope decreases (increases) upon expansion (compression), with one nodal point in between. In the resulting SN light curve, \citet{goldberg2020a} find an increased brightness at earlier times for an expanded progenitor. This is very similar to our findings. However, in our case, the density in the inner part of the envelope increases because of the restructuring after the cooling catastrophe. The resulting effect on the SN light curves is very similar because the density structure is modified similarly. In our case, we find more severe changes in the SN light curves because the envelope restructuring together with the large amplitude pulsations introduce a larger deviation from the hydrostatic structure compared to the effect of the first-overtone pulsations reported in \citet{goldberg2020a}  

In general, simulations of the core evolution and the dynamical envelope evolution at the same time to capture the onset of the pulsations and to simulate them to core collapse are not possible. The difference in the dynamical timescale of the envelope and the nuclear timescale in the core is too large. Nonetheless, it is important to start simulating the pulsations early enough before the core collapses, such that they can fully develop and possibly reach an equilibrium state, because we find that their effect on the final stellar structure and explosion properties is significant.

\subsection{Convection in pulsating envelope}\label{sec:discussion:convection}
The envelopes of RSGs are unstable against convection (\ie, the bulk motion of matter). The direct interplay between convection and pulsation is not well understood \citep[\eg][]{wood2007a,houdek2015a,freytag2017a,goldberg2022a,ahmad2023a,ma2024a}. Convection itself is a multidimensional phenomenon that might trigger non-radial pulsations, especially given that only a few convective cells are expected to be present in the envelope of RSGs \citep{goldberg2022a,ahmad2025a}. These can remove energy from the radial pulsations and reduce their amplitudes, weakening the overall effects of the pulsations and potentially smoothening the RSG light curve (Fig.~\ref{fig:dusty_LC}). \citet{olivier2005a,olivier2006a} include turbulent viscosity when modeling pulsations of red giant stars. This introduces damping of the pulsations and reduces their amplitudes. In our models, we find that increasing the mixing-length parameter of convection has a similar effect (see~Appendix~\ref{app:alpha_MLT}). However, the physical mechanisms at work are different because the increased mixing length leads to more efficient energy transport and more compact RSGs. There also exist alternative treatments to classical mixing-length theory \citep{prandtl1925a,cox1968a}, such as time-dependent models for convection \citep[\eg][]{kuhfuss1986a,xiong1997a}, which might offer a better description of convection in stellar envelopes. Including such models in simulations of oscillating red giants, \citet{xiong1997a} concluded that turbulent pressure can have an important effect on the pulsations. However, it has been shown by different studies that modifying the treatment of convection has negligible effects on the pulsations, that is, by introducing a phase shift between the pulsations and the convective flux \citep{langer1971a} or completely freezing the convective flux \citep{heger1997a}. In general, three-dimensional simulations of convective envelopes \citep[\eg][]{goldberg2022a,freytag2023a,ma2024a} can help us to better constrain the interplay between convection and pulsations.

Recent observations of asymptotic-giant-branch stars have revealed that the size of the convective cells changes throughout the pulsation cycle \citep{rosales-guzman2024a}. This might open up a new pathway to directly study the connection between convection and pulsations without the need for numerical simulations.

The treatment and efficiency of convection can directly impact the amplitude and period of the pulsations. This study aims to show that the pre-SN pulsations of RSGs exist and that they have a direct impact on the SN light curve. The role of convection can weaken the results on the SN light curves via lower amplitude pulsations. Nonetheless, we expect the amplitude to increase with the mass of the RSG; therefore, we still expect to see similar effects in higher-mass RSGs.

\subsection{Limitation of pulsations models}\label{sec:discussion:lims}
There are several limitations to the models presented above. They mostly originate from the general limitation of 1D models compared to 3D models.

The envelopes of RSGs are dominated by large convective cells \citep{goldberg2022a}, which cannot be treated in 1D (see discussion above; Sect.~\ref{sec:discussion:convection}). By assuming spherical symmetry for 1D simulations, any inhomogeneities in the envelope that can be induced by the large-scale convective motion in 3D are removed. Full 3D radiation-hydrodynamical simulations \citep[\eg][]{goldberg2022a, ahmad2023a} show that these variations can be significant, especially in the outermost layers of the stars. Consequently, the energy transport by radiation through these layers is changed and can be more efficient than expected by 1D models with standard MLT treatment \citet{goldberg2022a}. This might further have implications for the expected wind mass loss, which could be enhanced in certain locations of the surface, \eg, similar to the great dimming of Betelgeuse \citep{dupree2022a}. All of these effects can decrease the strength of the pulsations and the resulting diversity in supernova properties. The non-sphericity can immediately impact the early evolution of the supernovae light curve; the shock breakout will happen at different radii and at different times, lowering the overall luminosity during this phase \citep{goldberg2022b}.

The structure of the 1D envelope of the RSG shows several prominent density inversions (see Fig.~\ref{fig:density_profiles}). In a 3D realization of such an envelope, one might suspect that these inversions become weaker because they are Rayleigh-Taylor unstable. However, the dynamical timescale is similar to the evolutionary timescales of these features, making it unclear whether there is enough time to completely dissolve them in the envelope structure. Additionally, these density inversions are direct consequences of the recombination processes in the envelope. The released recombination energy during the expansion can continue to drive and maintain such density inversions during the period of the recombination. Any of these effects can result in the reduction of the pulsation amplitudes and reduce the diversity found in the supernova properties. These might also smooth the progenitor light curve, removing any short-duration peaks and producing a more sinusoidal light curve (see Fig.~\ref{fig:dusty_LC}).

Another limitation is that we need to assume a mixing length parameter $\alpha_\mathrm{MLT}$ for the treatment of convection. For our fiducial model, we assume $\alpha_\mathrm{MLT}=1.8$, which is inspired by calibrations to the Sun but also reproduces the HRD locations of Galactic RSGs \citep[\eg][]{ekstrom2012a}. A comparison between observed supernova properties and the predictions from RSG progenitor models led to the suggestion of using a larger mixing length parameter ($\alpha_\mathrm{MLT}\approx 3$) that produces more compact RSG progenitors \citep[\eg,][]{dessart2013a,paxton2018a}. Three-dimensional radiation-hydrodynamical simulations of convection in RSGs also support $\alpha_\mathrm{MLT}\approx 3-4$ deep in the envelope, but find lower $\alpha_\mathrm{MLT}$ near the surface (\citealp{goldberg2022a}, J.-Z. Ma et al., in prep.). The usage of \texttt{MESA}'s MLT++ in our simulation reduces the superadiabacity in radiation-dominated convective layers \citep{paxton2013a}, making convection more efficient, as suggested by the 3D simulations.

Progenitors with more efficient convection (\eg, $\alpha_\mathrm{MLT}=3$) show at a fixed mass lower amplitude pulsations and less diversity in the supernova light curves (see Appendix~\ref{app:alpha_MLT}). However, when adopting a larger $\alpha_\mathrm{MLT}$, more massive RSGs are still expected to exhibit larger amplitude pulsations. Although the radii of RSGs with a higher $\alpha_\mathrm{MLT}$ are generally smaller, they increase with increasing stellar mass, meaning that also the pulsation amplitudes increase. A more massive RSG with a higher $\alpha_\mathrm{MLT}$ is expected to show pulsations of similar amplitude as a lower mass RSG with a lower $\alpha_\mathrm{MLT}$. Therefore, increasing the mixing length parameter has the effect of increasing the initial mass of stars at which we expect large-amplitude pulsations to be present.

\subsection{Inferring explosion and pre-supernova properties}
There exist numerous techniques to estimate explosion and progenitor properties of Type~II SNe from the light curve properties \citep[\eg][]{popov1993a,goldberg2019a}. We show in Sect.~\ref{sec:LC} that the explosion of the same star with identical explosion properties but taken at different pulsation phases can lead to significantly different SN light curves. Because the density structure is non-trivially modified by the pulsations compared to the hydrostatic structure, scaling relations fail to connect the SN progenitor properties to the SN light curve \citep{goldberg2020a}. Therefore, inferences of SN progenitor properties from the SN light curve need to be treated with care when the progenitor star is pulsating, especially when the amplitude of the pulsation is large. 

Moreover, inferring progenitor properties based on archival imaging can lead to incorrect luminosity and temperature estimates, especially when only one epoch of observations is available. The uncertainties we expect from the trace of the RSG in the HRD during the pulsations can be even larger than observational uncertainties (see Fig.~\ref{fig:pulsations_HRD}c). Therefore, observing the progenitor at just one instance of time will lead to a significant over-/underestimate of its luminosity and/or temperature. Future surveys, such as the Legacy Survey of Space and Time (LSST) at the Vera C.\ Rubin Observatory \citep{ivezic2019a}, can increase the number of progenitors and may provide even full progenitor light curves.

\section{Conclusion}\label{sec:conclusion}
In this paper, we study the pulsations of RSGs and their immediate implications on the light curves of Type~II SNe. We follow the final evolutionary stages of a \qty{15}{\msun} RSG. Our results can be summarized as follows:
\begin{itemize}
    \item By direct hydrodynamical simulations of the red supergiant, we find radial pulsations during core carbon burning. During core helium burning, the RSG envelope remains stable. These findings confirm earlier work by \citet{heger1997a}, \citet{yoon2010a}, and \citet{clayton2018a}.
    \item We show that these pulsations are driven by the $\kappa\gamma-$mechanism through the cyclic recombination and ionization of hydrogen and singly- and doubly-ionized helium.
    \item The pulsation period agrees with the period-luminosity relation of observed RSG. The amplitude of the brightness variations is larger compared to observations. We predict that such large-amplitude variations are rare because they only occur shortly before core collapse.
    \item We find that taking the radial pulsations of massive red supergiants into account leads to light curves that are significantly different from light curves obtained for a hydrostatic model. 
    \item The light curves at different pulsation phases naturally reproduce some of the diversity of hydrogen-rich supernovae light curves. We find different decline rates in the supernova light curve, with the fastest-declining light curves for the most extended progenitors and slow-declining light curves from the more compact progenitors. The light curves also show early excess emission without the need to invoke extra CSM around the supernova.
    \item There is a clear distinction between the photospheric velocity evolution of the fast and slow-declining SNe. For the fast-declining (Type~II-L) SNe, the photospheric velocities plateau for up to \qty{30}{d} and reach the smallest maximum velocities. Slow declining (Type~II-P) SNe reach the larger maximum photospheric velocities that monotonously decrease afterwards.
    \item We find that the light curve shapes and features can be connected to the pre-supernova density profiles, which are directly impacted by the locations of the partial-ionization zones.
    \item The radial pulsations before the SN explosion cause variations in the luminosity and effective temperature of the RSG by up to an order of magnitude. Hence, the locations of massive red supergiants in the HRD undergo significant changes before core collapse, implying that only limited constraints can be gained from the observation of pre-supernova luminosities of red supergiants. This uncertainty is expected to increase for more massive progenitors, where we even expect dynamical mass ejection \citep{clayton2018a}, directly impacting the missing red-supergiant problem.
    
\end{itemize}

We aim to extend this study by analyzing a larger parameter space of RGSs, \eg, by studying RSGs with different initial masses. This way, we can learn about the pulsations with different amplitudes and their effect on the SN light curves \citepalias{paperII}.

\begin{acknowledgements}
    We thank J.~Spyromilio for interesting discussions at the MIAPbP 2023 (Interacting Supernovae) that triggered this work. We thank the referee for useful and constructive feedback that helped to improve the paper. We thank J.~Goldberg and E.~Farag for constructive discussions that helped improve the quality of this work. VAB, EL, FRNS, and PhP acknowledge support from the Klaus Tschira Foundation.
    This work has received funding from the European Research Council (ERC) under the European Union’s Horizon 2020 research and innovation programme (Grant agreement No.\ 945806), and is supported by the Deutsche Forschungsgemeinschaft (DFG, German Research Foundation) under Germany’s Excellence Strategy EXC 2181/1-390900948 (the Heidelberg STRUCTURES Excellence Cluster). VAB acknowledges support from the International Max Planck Research School for Astronomy and Cosmic Physics at the University of Heidelberg (IMPRS-HD). EL acknowledges support through a start-up grant from the Internal Funds KU Leuven (STG/24/073) and through a Veni grant (VI.Veni.232.205) from the Netherlands Organization for Scientific Research (NWO). We acknowledge with thanks the variable star observations from the AAVSO International Database contributed by observers worldwide and used in this research. This research was supported by the Munich Institute for Astro-, Particle and BioPhysics (MIAPbP) which is funded by the Deutsche Forschungsgemeinschaft (DFG, German Research Foundation) under Germany´s Excellence Strategy – EXC-2094 – 390783311.\\
    \textbf{Software:} \texttt{astropy} \citep{astropycollaboration2013a,astropycollaboration2018a,astropycollaboration2022a}, \texttt{CMasher} \citep{velden2020a}, \texttt{matplotlib} \citep{hunter2007a}, \texttt{numpy}, \citep{harris2020a}, \texttt{PyAstronomy} \citep{czesla2019a}, \texttt{PYPHOT} \citep{fouesneau2025a}, \texttt{scipy} \citep{virtanen2020a}, \texttt{TULIPS} \citep{laplace2022b}.
\end{acknowledgements}

\bibliographystyle{aa}
\bibliography{refs_paper}

\begin{appendix}

\section{Damping the pulsations}\label{app:damping}

When initializing the hydrodynamical simulation, spurious velocities might be introduced that can dominate the hydrodynamical evolution of the model. To test whether this is the case, we apply an initial damping force that is slowly reduced over time. This is achieved by extending the momentum equation via   
\begin{equation}
    \dot{v} = \gamma v,
\end{equation}
with a damping factor $\gamma$. This new term causes an exponential decay of any velocities on a timescale of $1/\gamma$. We follow the description of \citet{pakmor2012b} and \citet{ohlmann2016c} and choose an initial damping timescale of $\tau_\mathrm{dyn}/10$ that is then exponentially increased to $\tau_\mathrm{dyn}$, prescribed by 
\begin{equation}
    \gamma(t) = \left\{
    \begin{array}{c@{\quad}l}
        \dfrac{10}{\tau_\mathrm{dyn}} & t < 10\tau_\mathrm{dyn} \\ & \\
        \dfrac{10}{\tau_\mathrm{dyn} \times 10^\frac{t - 10\tau_\mathrm{dyn}}{10\tau_\mathrm{dyn}}} &  10\tau_\mathrm{dyn}\leq t \leq 20\tau_\mathrm{dyn}. \\ & \\
        0 & t > 20\tau_\mathrm{dyn}
    \end{array}
    \right.
\end{equation}
This ensures that any pulsations caused by the changing damping factor have enough time to decay. After $20\tau_\mathrm{dyn}$, the damping is turned off and the model is allowed to evolve freely.

\begin{figure}[h!]
    \centering
    \resizebox{\hsize}{!}{\includegraphics{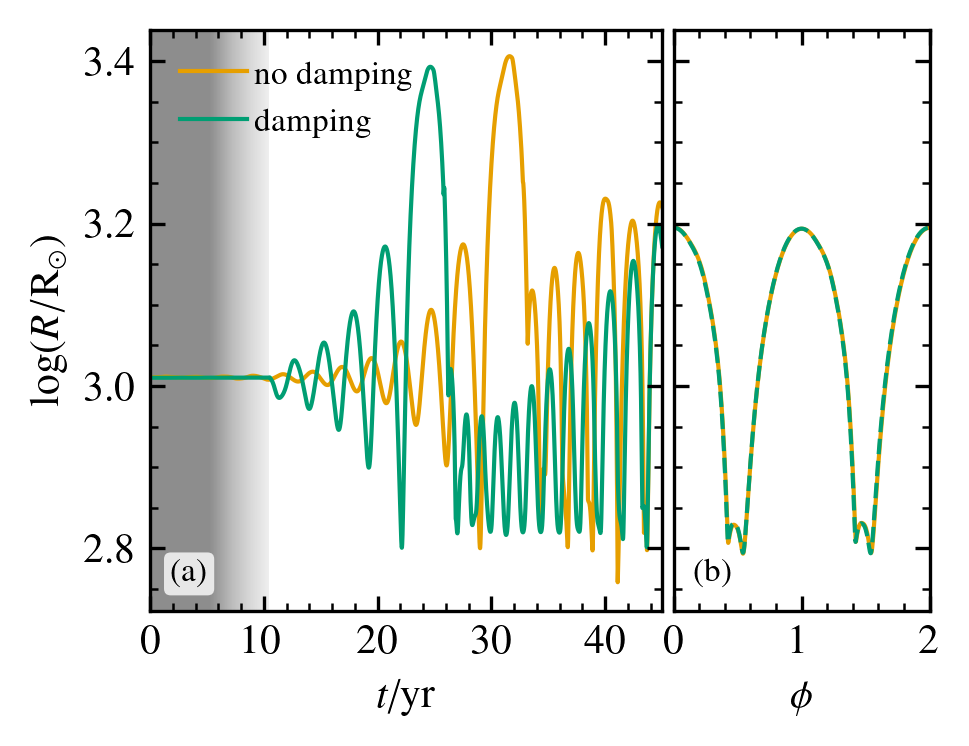}}
    \caption{Comparison of the growth of the pulsations (panel a) and the steady state pulsations (panel b) between the undamped and damped model. The shading during the initial $\sim\qty{10}{yr}$ indicates the magnitude of the damping factor $\gamma$.}
    \label{fig:damping}
\end{figure}

When applying this damping scheme to the $15\,\msun$ RSG model, we find that the pulsations are quenched for about $10\,\mathrm{yr}$ or about $20 \tau_\mathrm{dyn}$ (Fig.~\ref{fig:damping}). Then, the model starts to pulsate and experiences a catastrophic cooling event that is accompanied by restructuring of the envelope. Afterwards, we find steady pulsations with an identical pulsation period of $817^{+8}_{-10}\,\mathrm{d}$ and the radius evolution differs by less than \qty{0.4}{\%}. 

\begin{figure}
    \centering
    \resizebox{\hsize}{!}{\includegraphics{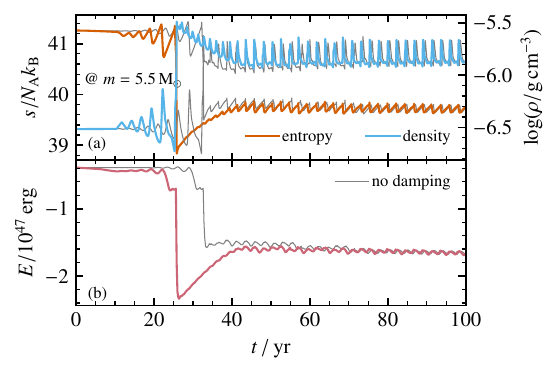}}
    \caption{Restructuring of the envelope for the model with initial damping, similar to Fig.~\ref{fig:restructuring}. Panel~(a) shows the specific entropy in units of Avogadro's number $N_\mathrm{A}$ and the Boltzmann constant $k_\mathrm{B}$, as well as the density. Both quantities are reported at a mass coordinate of \qty{5.5}{\msun}, which is close to the bottom of the convective envelope. The total energy of the convective envelope is shown in panel (b). For comparison, the undamped model is shown as well (same as Fig.~\ref{fig:restructuring}c and e).}
    \label{fig:damping_restruct}
\end{figure}

Once the damping is turned off after about $ 10\,\mathrm{yr}$, the envelope starts to contract immediately. Afterwards, the pulsations start at a much larger amplitude compared to the undamped case, and the catastrophic cooling is already reached after about $24\,\mathrm{yr}$ (Fig.~\ref{fig:damping_restruct}). The reason for the more violent initial evolution of the model with damping is that the thermal timescale of the envelope ($\sim 11\,\mathrm{yr}$) is similar to the damping duration. This means that any thermal timescale evolution of the envelope, \eg, caused by the net energy loss at the surface via radiation, cannot be compensated for by contraction or expansion (Fig.~\ref{fig:damping_restruct}b). Only once the damping term is small enough or turned off entirely does the envelope adjust. In our case, this means that the envelope contracts on the dynamical timescales and initiates the large-amplitude initial pulsation. Nonetheless, we find a growth rate of $\eta\approx 1.9$, which is very close to the undamped case. This indicates that even though the start of the pulsations is more violent, the growth is identical to the undamped case.

During the catastrophic cooling event, we find that more energy of the envelope is radiated away compared to the undamped case. This results in a different envelope structure immediately after the catastrophic cooling. However, at $t\approx\qty{70}{yr}$, both models converge to the same structure on the thermal timescale (Fig.~\ref{fig:damping_restruct}). This shows that even though the evolution and the growth of the pulsations between the two models differ, they reach the same envelope structure after the catastrophic cooling and the accompanied restructuring. 

Generally, the damping has negligible effects on the steady-state pulsations that we are interested in and which can be present for a sustained duration before the supernova. This shows that these pulsations are robust, even if the initial evolution and growth of the pulsations are altered.  

\FloatBarrier

\section{Different core masses}\label{app:Mcut}
\begin{figure}[h!]
    \centering
    \resizebox{\hsize}{!}{\includegraphics{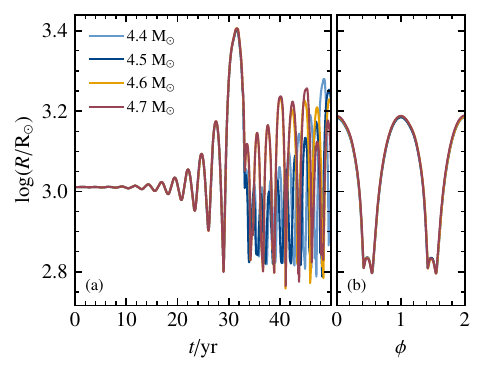}}
    \caption{Comparison of the growth of the pulsations (panel a) and the steady state pulsations (panel b) between the model with different cut-out core masses. The total helium core mass is $5.1\,\msun$, defined by a hydrogen abundance less than 0.1.}
    \label{fig:Mcut}
\end{figure}

For our fiducial model of the $15\,\msun$ RSG, we remove the core at a mass coordinate of $4.6\,\msun$. To test whether this choice has a significant impact on the steady-state pulsations, we compute models with a removed core of $4.4$, $4.5$, and $4.7\,\msun$. The growth of the pulsations up to the catastrophic cooling event and the envelope restructuring are identical between all the models (Fig.~\ref{fig:Mcut}). We find differences afterwards when the models adjust to the new structure and the pulsations are not yet periodic. However, once the steady-state pulsations are reached, we find that the pulsation periods differ by less than \qty{1}{\%} and the radius evolution differs by less than \qty{0.5}{\%}. Therefore, we conclude that the exact location of the core removal has only a minor effect on the steady-state pulsations of the RSG envelope.

\FloatBarrier

\section{Dependence of the pulsations on the convective mixing length parameter}\label{app:alpha_MLT}
\begin{figure}
    \centering
    \resizebox{\hsize}{!}{\includegraphics{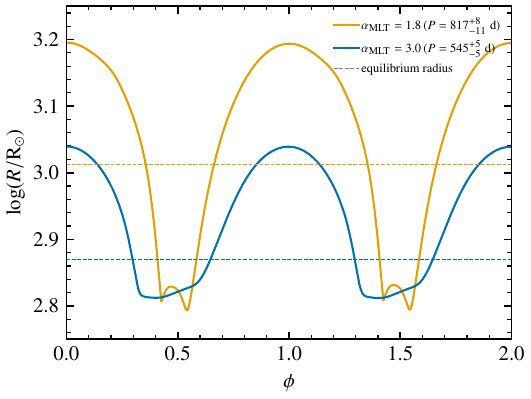}}
    \caption{Comparison of the pulsations of two RSG models with different $\alpha_\mathrm{MLT}$ values over two pulsation cycles.}
    \label{fig:alpha_MLT}
\end{figure}

\begin{figure}
    \centering
    \resizebox{\hsize}{!}{\includegraphics{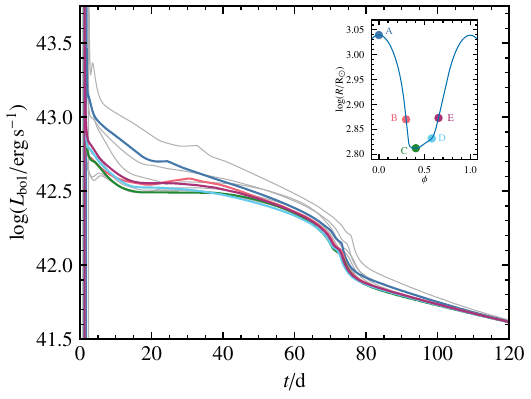}}
    \caption{Similar to Fig.~\ref{fig:LC_vphot}a, but for the progenitor with $\alpha_\mathrm{MLT}=3$, where different colors correspond to different explosion phases. The light curves for the progenitor with $\alpha_\mathrm{MLT}=1.8$ are shown with gray colors as a comparison.}
    \label{fig:alpha_MLT_LCs}
\end{figure}

In Fig.~\ref{fig:alpha_MLT}, we show a comparison of the pulsations of the \qty{15}{\msun} RSG model (i.e., mixing length parameter $\alpha_\mathrm{MLT} = 1.8$) and a \qty{15}{\msun} RSG with $\alpha_\mathrm{MLT} = 3.0$. For this, we take the initial stellar structure at the end of core carbon burning and adjust $\alpha_\mathrm{MLT}$ to $3.0$ in the envelope only. In particular, this means that the core structure is unchanged and $L/M$ is the same, ensuring a meaningful comparison between the models. The larger $\alpha_\mathrm{MLT}$, the more efficient the energy transport via convection. The computational methods are identical to the description in Sects.~\ref{sec:methods:SE} and \ref{sec:methods:pulsations}. The radius of the RSG with the increased value of $\alpha_\mathrm{MLT}$ is \qty{740}{\rsun} compared to \qty{1024}{\rsun} with the lower value of $\alpha_\mathrm{MLT}$. Moreover, we find both a shorter pulsation period ($545\pm5 \, \mathrm{d}$) and a lower pulsation amplitude for the high $\alpha_\mathrm{MLT}$ RSG. This shows that the treatment of convection directly modulates the strength of the pulsations. 

Additionally, the model with the larger $\alpha_\mathrm{MLT}$ avoids a catastrophic cooling event, because it stays generally more compact, which causes the thermal timescale ($\propto R^{-1}$) to be always much larger than the dynamical timescale ($\propto R^{3/2}$). Therefore, no significant envelope restructuring occurs in this model.

We compute supernova light curves at characteristic points during the pulsation cycle as described in Sect.~\ref{sec:methods:LC}. The explosion parameters are unchanged because the core structure is identical to the $\alpha_\mathrm{MLT}=1.8$ model. The light curves of the resulting supernovae are shown in Fig.~\ref{fig:alpha_MLT_LCs}. The diversity of the light curves because of the pulsating progenitor is weaker compared to the $\alpha_\mathrm{MLT}=1.8$ case. This is expected because the amplitude of the pulsation is lower, which causes weaker structural changes in the envelope. 

We note that the choice of $\alpha_\mathrm{MLT}$ in the convective envelope of RSGs directly impacts the mass loss because of the different radii. This can then lead to different envelope masses at core collapse and therefore also to different $L/M$. Because $L/M$ directly impacts the pulsation properties, we expect larger differences between the pulsations when $\alpha_\mathrm{MLT}=3$ is applied through the entire RSG evolution phase. 

\FloatBarrier

\section{Dust model interpolation}\label{app:DM_interp}
The interpolation of the \texttt{DUSTY}+MARCS (DM) models requires some extra details. First, all the MARCS models are down-scaled to a spectral resolution of 1000. Then, all the DM models are calculated at the reference luminosity of $L_0 = \qty{1}{\lsun}$ and are later rescaled to the nominal luminosity of the RSG. Calculating all DM models at the same luminosity has the advantage that the interpolation between the models can be done directly. We interpolate the absolute magnitudes $M_X$ in filters $V$ and $K$, and the inner dust radius $r_1$, because all are later needed for calculating the RSG light curve. Each DM model (\ie each combination of $T_\mathrm{eff}$ and $\log g$) consists of 31 dust models with varying amounts of dust, represented by the optical depth $\tau$ linearly increasing from $\log \tau = -1$ to $\log \tau = 2$. 

\begin{figure}
    \centering
    \resizebox{\hsize}{!}{\includegraphics{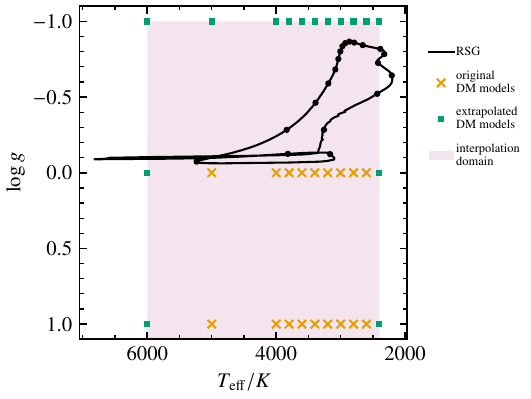}}
    \caption{Kiel diagram showing the RSG pulsation cycle and the MARCS model coverage. Dots along the RSG track are spaced equally in time every 1/20 of the pulsation period. The crosses show the combinations of $T_\mathrm{eff}$ and $\log g$ for which there are MARCS models available and for which we can compute \texttt{DUSTY}+MARCS (DM) models. After extending the grid by one grid spacing in each direction (green squares), we can cover \qty{80}{\%} of the RSG pulsation cycle by linearly interpolating the newly obtained grid of DM models (shaded region).}
    \label{fig:MARCS_interpolation}
\end{figure}

\begin{figure}
    \centering
    \resizebox{\hsize}{!}{\includegraphics{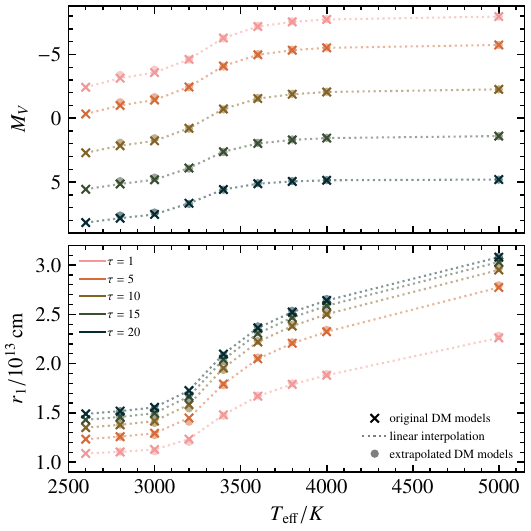}}
    \caption{Absolute $V$-band magnitude and the inner dust radius $r_1$ for the DM models with $\log g = 0.0$ as a function of effective temperature for various optical depths. For comparison, we extrapolate the DM models from $\log g = 1.0$ and 2.0 to $\log g=0.0$ wherever the models are available.}
    \label{fig:test_dust_interp}
\end{figure}

Next, we extrapolate the DM models by one grid spacing in $T_\mathrm{eff}$ at constant $\log g$, both at the low and at the high-temperature edge of the grid. Afterward, we extrapolate the DM models by one grid spacing in $\log g$ at constant $T_\mathrm{eff}$ towards the low surface gravity edge of the model grid. Fig.~\ref{fig:test_dust_interp} shows that the extrapolation along $\log g$ works well. Once we extend the grid, we can cover \qty{80}{\%} of the RSG pulsation cycle with the DM model grid (Fig.~\ref{fig:MARCS_interpolation}). The interpolation of the DM models is done linearly on the extended DM model grid.

To adjust the interpolated DM models to the nominal luminosity $L$ of the RSG, we need to rescale the inner dust radius $r_1$ and absolute magnitudes $M_X$ via
\begin{eqnarray}
    M_X &\rightarrow& M_X - 2.5 \log_{10}(L/L_0), \\
    r_1 &\rightarrow& r_1 \sqrt{L/L_0},
\end{eqnarray}
with the reference luminosity $L_0 = \qty{1}{\lsun}$ \citep{ivezic1997a}. 

To construct the light curve, we need to know the time evolution of the optical depth $\tau$. The optical depth is not expected to be constant because the pulsating RSG changes both in luminosity and effective temperature. Hence, the inner dust radius $r_1$, defined at a dust condensation temperature of \qty{800}{K}, needs to vary with time as well. First, we fix the optical depth $\tau_0$ at maximum extent ($\phi = 0$). For each DM model, we linearly interpolate along the $\tau$ axis to obtain the quantities at the desired optical depth. Based on $\tau_0$, we can determine the dust density $\rho_{x}$ at an arbitrary but fixed distance $x$, that satisfies $r_1 \leq x \leq 10^3 r_1$ for all DM models simultaneously, via
\begin{equation}\label{eq:rhox}
    \rho_{x} = \rho_1 \left(\frac{r_1}{x}\right)^2. 
\end{equation}
The density $\rho_1$ at distance $r_1$ is given by 
\begin{equation}
    \rho_1 = \frac{\tau \eta_1}{r_1 \mu/\rho} \propto \frac{\tau}{r_1},
\end{equation}
with $\eta_1$ the normalized dimensionless density at $r_1$ (see \citealt[equation~4]{ivezic1997a} for the definition of $\eta$) and $\mu/\rho$ the mass attenuation coefficient. Both $\eta_1$ and $\mu/\rho$ are constant amongst all DM models because we assume the same density scaling and dust composition. The power-law scaling of the density profile is set by the input physics of the \texttt{DUSTY} models (Sect.~\ref{sec:methods:dust}).

To obtain the $\tau$ evolution of the light curve, we require the density $\rho_x$ to be constant throughout the pulsation cycle. With this assumption and together with Eq.~\eqref{eq:rhox}, we can determine the optical depth $\tau$ for any DM model at given phase $\phi$. The absolute magnitudes $M_X$ are obtained by linear interpolation of the DM model along the $\tau$ axis. This way, $M_X$ can be determined for all the points during the pulsation cycle, allowing us to construct the entire light curve. 

\FloatBarrier

\section{Envelope structure during the pulsation cycle}\label{app:pulsations}
\begin{figure*}
    \centering
    \includegraphics[width=17cm]{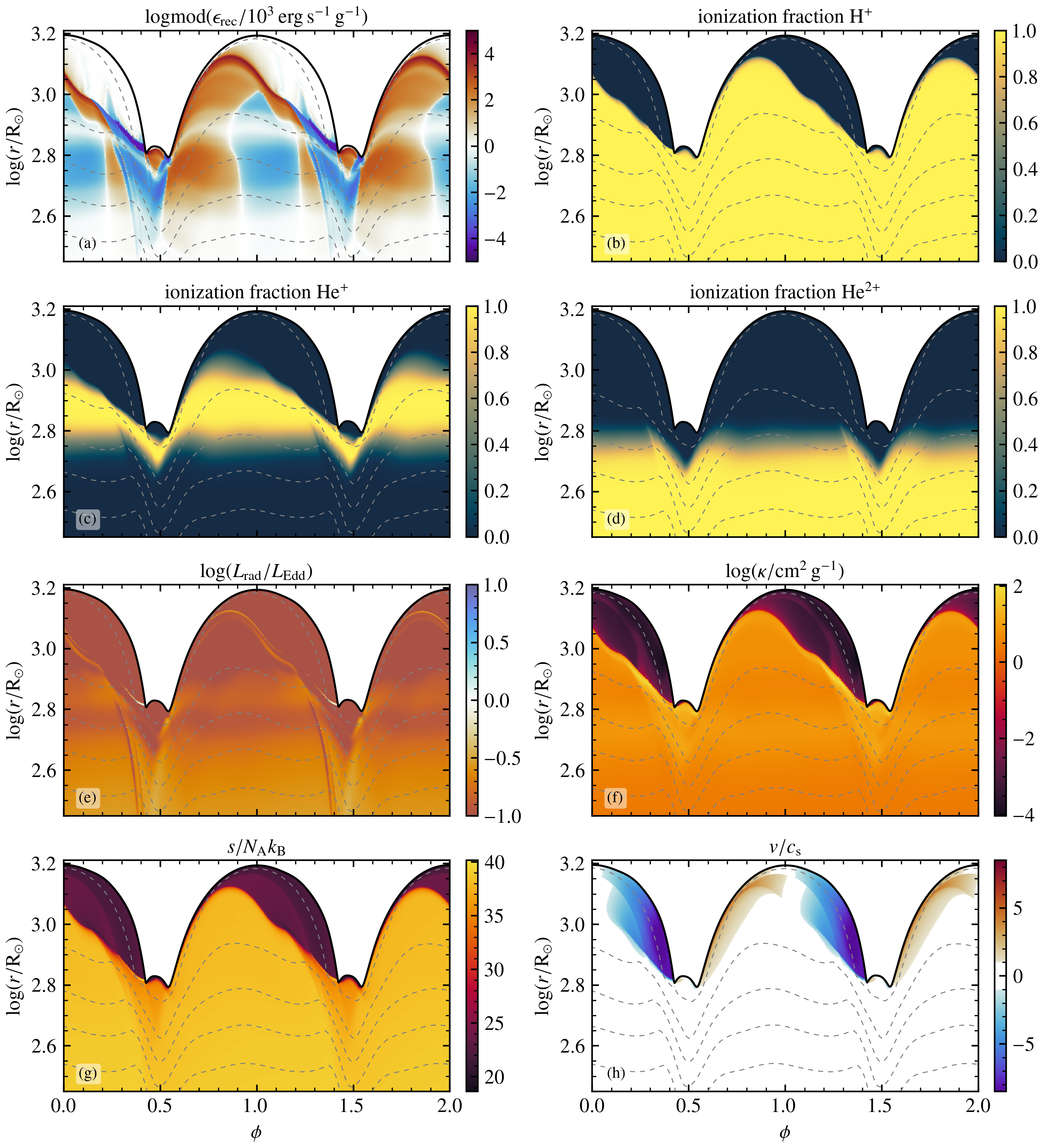}
    \caption{Structure evolution diagrams of the envelope of 2 full pulsation cycles. Colors show the specific released recombination energy $\epsilon_\mathrm{rec}$ in panel (a), the ionization fractions of hydrogen and helium in panels (b)-(d), the radiative luminosity $L_\mathrm{rad}$ in units of the Eddington luminosity $L_\mathrm{Edd}$ in panel (e), the opacity $\kappa$ in panel (f), the specific entropy $s$ in units of Avogadro's number $N_\mathrm{A}$ and the Boltzmann constant $k_\mathrm{B}$ in panel (g), and the radial velocity $v$ in units of the sound speed $c_\mathrm{s}$ in panel (h). The dashed lines represent lines of constant mass for masses of 12, 11, 10, 9, and \qty{8}{\msun} from top to bottom. For better visualization, the logmod transformation is used for the specific recombination energy, with $\mathrm{logmod}(x) = \mathrm{sign}(x)\log_{10}(|x| + 1)$ \citep{john1980a}.}
    \label{fig:pulsations_kipp1}
\end{figure*}

\begin{figure*}
    \centering
    \includegraphics[width=17cm]{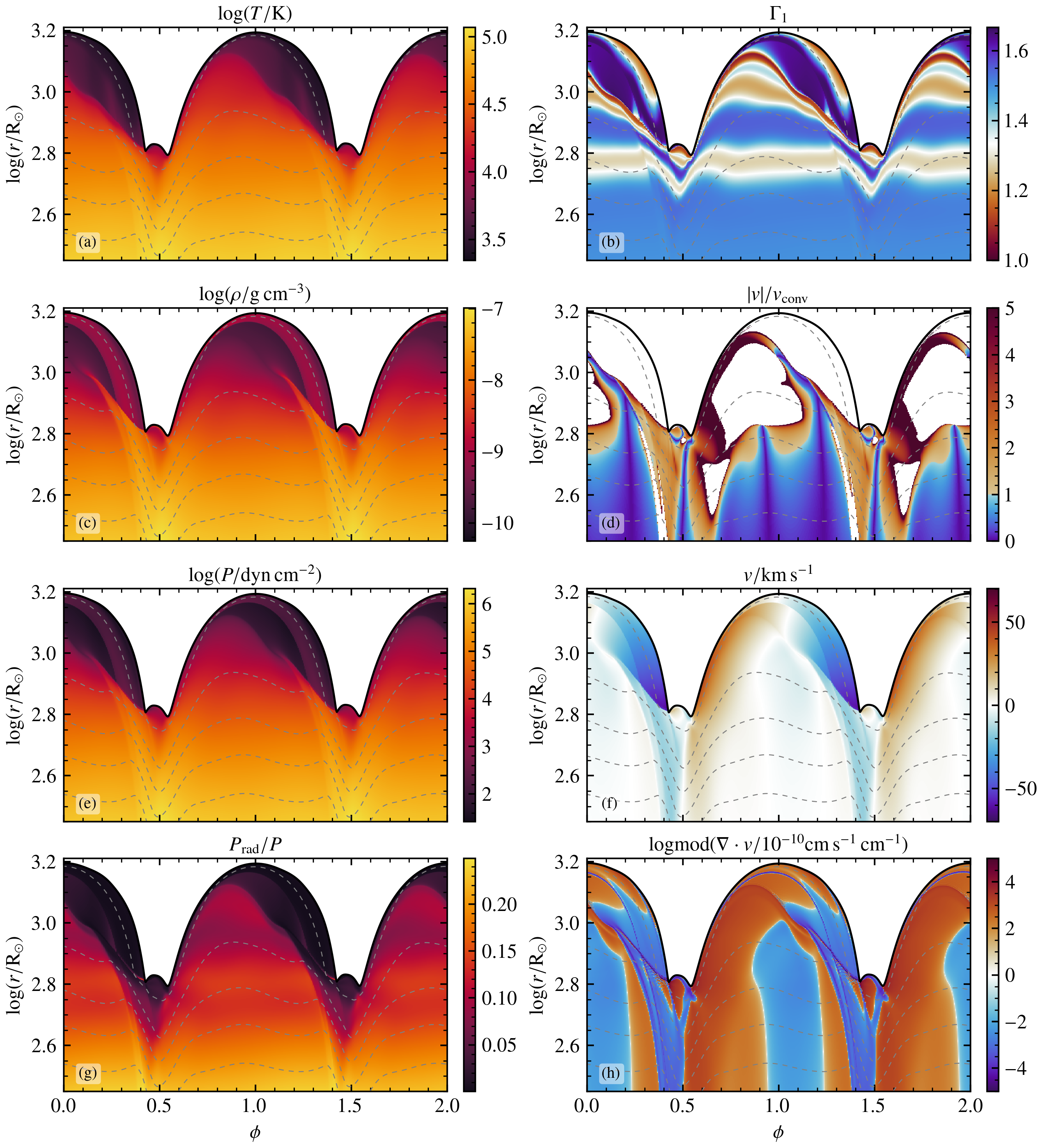}
    \caption{Structure evolution diagrams similar to Fig.~\ref{fig:pulsations_kipp1}. Colors show the temperature $T$ in panel (a), the first adiabatic index $\Gamma_1 = (\partial \ln P/ \partial \ln \rho)_{s}$ in panel (b), the density $\rho$ in panel (c), the absolute radial velocity in units of the convective velocity $v_\mathrm{conv}$ in panel (d), the total pressure $P$ in panel (e), the radial velocity in panel (f), the fraction of the radiation pressure $P_\mathrm{rad}$ of the total pressure in panel (g), and the velocity divergence $\nabla \cdot v$ in panel (h).} 
    \label{fig:pulsations_kipp2}
\end{figure*}

We show various structure evolution diagrams of the envelope throughout the pulsation cycle in Figs.~\ref{fig:pulsations_kipp1} and~\ref{fig:pulsations_kipp2}. 

From the released recombination energy (Fig~\ref{fig:pulsations_kipp1}a), one can see the different phases of recombination and re-ionization of hydrogen and helium (see also Figs~\ref{fig:pulsations_kipp1}b-d). We find a short phase of recombination of He$^{2+}$ at $\phi = 0.3$ that leads to the expansion of intermediate layers. However, this expansion phase is quenched as the entire envelope above falls back. It seems that the natural oscillation frequency of an envelope driven by He$^{2+}$ recombination is shorter than the pulsation period we find for the entire RSG envelope.

The neutral envelope falls back at highly supersonic speeds (Fig.~\ref{fig:pulsations_kipp1}h) and is re-ionized at very large rates ($\approx \qty{10^8}{erg \, s^{-1} \, g^{-1}}$). This causes the formation of shock waves (see Fig.~\ref{fig:pulsations_kipp2}h that traces the shocks via the velocity divergence). In these layers, the opacity reaches the highest values throughout the envelope (Fig.~\ref{fig:pulsations_kipp1}f). This also causes the radiative luminosity to locally exceed the Eddington luminosity (Fig.~\ref{fig:pulsations_kipp1}). The re-ionization layer causes a sharp increase in density (Fig.~\ref{fig:pulsations_kipp2}c) and pressure (Fig.~\ref{fig:pulsations_kipp2}e). Ultimately, all these factors lead to the bounce-off and re-expansion of the outer layers, assisted by recombination. The constant-mass lines in Fig.~\ref{fig:pulsations_kipp1} clearly show that the layers further inside the envelope keep contracting. As they become more compressed, the temperature (Fig.~\ref{fig:pulsations_kipp2}a) and the pressure increase, slowing down the layer's contraction and starting its re-expansion at $\phi=0.5$. This turnaround process deep inside the envelope is much more gradual compared to the outer layers.

Around $\phi=0.7$, the outer layers of the envelope expand at supersonic velocities (Fig.~\ref{fig:pulsations_kipp1}h). In some of these layers, energy transport is moderated via convection. We find that the expansion velocity of these layers exceeds the convective velocity that is predicted by mixing length theory (Fig~\ref{fig:pulsations_kipp2}d). The application of mixing length theory to these layers is most certainly not appropriate, as we deviate far from hydrostatic equilibrium and the simulation timescale is comparable to the convective turnover timescale.

\FloatBarrier

\section{Photosphere evolution}\label{app:SN_phot}
\begin{figure}[h!]
    \centering
    \resizebox{\hsize}{!}{\includegraphics{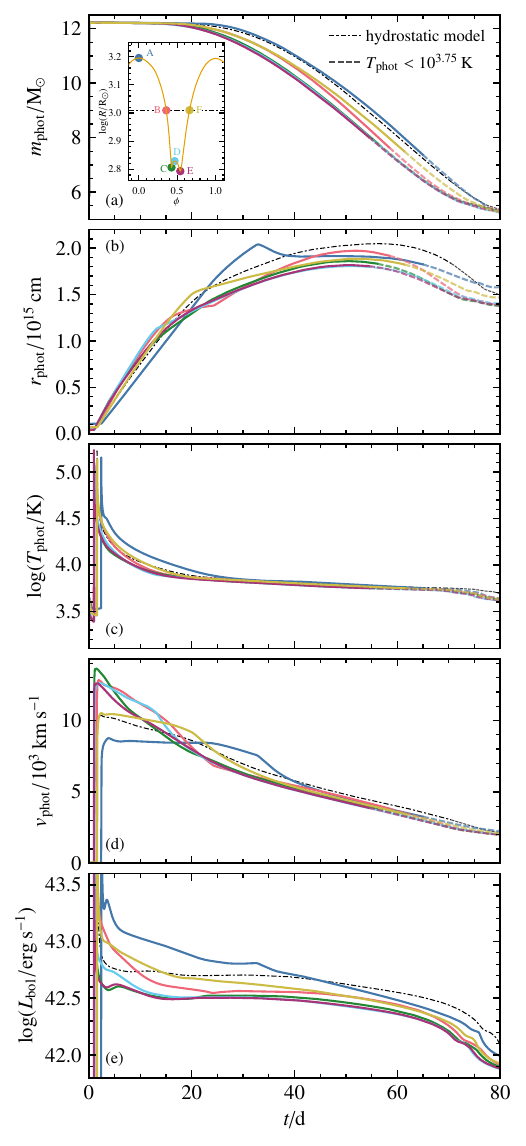}}
    \caption{Time evolution of the photosphere mass coordinate (panel a), the radius of the photosphere (panel b), the effective temperature of the photosphere (panel c), and the photospheric velocity (panel d). The light dashed lines indicate the phase when the photosphere temperature drops below $10^{3.75}\,\mathrm{K}$. At this point, \texttt{SNEC} has difficulties estimating the exact location of the photosphere \citep{morozova2015a}. The dash-dotted line illustrates the hydrostatic progenitor. The inset shows the radius evolution of one pulsation cycle and indicates the points A--F.}
    \label{fig:phot_props}
\end{figure}

We show the time evolution of the photosphere mass coordinate $m_\mathrm{phot}$, the photosphere radius $r_\mathrm{phot}$, and the effective temperature $T_\mathrm{phot}$ in Fig.~\ref{fig:phot_props}. For comparison, we also show the bolometric light curves and the photospheric velocities (see~Fig.~\ref{fig:LC_vphot}).

For $\qty{20}{d} < t < \qty{80}{d}$, we see a clear ordering of $m_\mathrm{phot}$ for any fixed time. For a larger progenitor radius, we find a larger photosphere mass coordinate. This means that the photosphere recedes later for the more extended progenitors (points A, B, and F) compared to the compact progenitors (points C, D, and E). The rate at which the photosphere moves to lower masses is similar. 

The photosphere radius of the most extended progenitor (point~A) is the smallest until $\sim \qty{30}{d}$, but then expands to the largest overall radius. This is a direct consequence of the flat photospheric velocity evolution and a lower peak velocity compared to the more compact progenitors. 

The supernova from the most extended progenitors also shows the highest effective temperatures during the first \qty{30}{d}. This is expected from the Stefan-Boltzmann law. The small photospheric radius and the large bolometric luminosity necessarily lead to a high effective temperature.

\FloatBarrier

\end{appendix}

\end{document}